%% file: main.tex
\documentclass[11pt]{article}

\input{texs/preamble}

\title{Trends in Frontier AI Model Count: A Forecast to 2028}
\author[1]{Iyngkarran Kumar\thanks{Lead author. Work completed as a Winter Fellow at the Centre for the Governance of AI. Send emails to iyngkarrankumar@gmail.com}}
\author[2]{Sam Manning}
\affil[1]{University of Edinburgh}
\affil[2]{Centre for the Governance of AI}

\date{\today}

\begin{document}

\maketitle

\input{texs/0_abstract}

\input{texs/1_introduction}
\input{texs/2_methodology}
\input{texs/3_results}
\input{texs/4_verification}
\input{texs/5_limitations}
\input{texs/6_discussion}
\input{texs/7_conclusion}
\input{texs/11_acknowledgments}
\printbibliography
\input{texs/8_appendix_a_c}
\input{texs/9_appendix_d_f}
\input{texs/10_appendix_g_i}

\end{document}

%% file: texs/preamble.tex
\usepackage{amsmath,amssymb}
\usepackage{graphicx}
\usepackage{hyperref}
\usepackage{caption}
\usepackage{subcaption}
\usepackage{booktabs}
\usepackage{multirow}
\usepackage{array}
\usepackage[top=1in, bottom=1in, left=1in, right=1in]{geometry}
\usepackage{xcolor}
\usepackage{float}
\usepackage{enumitem}
\usepackage{mathtools}
\usepackage{microtype}
\usepackage{authblk} %
\usepackage{comment}

\usepackage[backend=bibtex,style=authoryear,sorting=none]{biblatex}
\DeclareCiteCommand{\cite}
  {\usebibmacro{prenote}}
  {\usebibmacro{citeindex}%
   \printtext[brackets]{%
     \usebibmacro{cite}}}
  {\multicitedelim}
  {\usebibmacro{postnote}}

\DeclareCiteCommand*{\cite}
  {\usebibmacro{prenote}}
  {\usebibmacro{citeindex}%
   \printtext[brackets]{%
     \usebibmacro{citeyear}}}
  {\multicitedelim}
  {\usebibmacro{postnote}}

\DeclareCiteCommand{\parencite}[\mkbibparens]
  {\usebibmacro{prenote}}
  {\usebibmacro{citeindex}%
   \printtext[brackets]{%
     \usebibmacro{cite}}}
  {\multicitedelim}
  {\usebibmacro{postnote}}

\DeclareCiteCommand*{\parencite}[\mkbibparens]
  {\usebibmacro{prenote}}
  {\usebibmacro{citeindex}%
   \printtext[brackets]{%
     \usebibmacro{citeyear}}}
  {\multicitedelim}
  {\usebibmacro{postnote}}

%% file: texs/0_abstract.tex
\begin{abstract}
Governments are starting to impose requirements on AI models based on how much compute was used to train them. For example, the EU AI Act imposes requirements on providers of general-purpose AI with systemic risk, which includes systems trained using greater than $10^{25}$ floating point operations (FLOP). In the United States' AI Diffusion Framework, a training compute threshold of $10^{26}$ FLOP is used to identify ``controlled models'' which face a number of requirements. We explore how many models such training compute thresholds will capture over time. We estimate that by the end of 2028, there will be between 103-306 foundation models exceeding the $10^{25}$ FLOP threshold put forward in the EU AI Act (90\% CI), and 45-148 models exceeding the $10^{26}$ FLOP threshold that defines controlled models in the AI Diffusion Framework (90\% CI). We also find that the number of models exceeding these absolute compute thresholds each year will increase superlinearly -- that is, each successive year will see more new models captured within the threshold than the year before. Thresholds that are defined with respect to the largest training run to date (for example, such that all models within one order of magnitude of the largest training run to date are captured by the threshold) see a more stable trend, with a median forecast of 14-16 models being captured by this definition annually from 2025-2028.
\end{abstract}

%% file: texs/1_introduction.tex
\section{Introduction}

Recent years in machine learning have seen the rise of foundation models -- AI systems that exhibit powerful and general-purpose capabilities. Governments across the world are starting to impose requirements on the development and deployment of the most capable such systems, such as the GPT o-series \cite{openai_o1}. In December 2023 the European Union adopted the EU AI Act \cite{eu_ai_act}, the world's first comprehensive legislation designed to govern the development and use of AI systems. Among other things, the Act imposes requirements on providers of general-purpose AI with systemic risk (GPAISR), as of August 2025, which includes systems trained using greater than $10^{25}$ floating point operations (FLOP)\footnote{In this paper we often use the notation $10^{X}$ to refer to absolute compute thresholds. However the notation 1eX also appears at times which is interchangeable with the first notation. For example, 1e24 FLOP = $10^{24}$ FLOP, 5.3e25 FLOP = $5.3 \times 10^{25}$ FLOP, etc.}. The European Union is not the only jurisdiction to propose requirements based on training compute thresholds; in one of its final acts the Biden administration issued the Artificial Intelligence Diffusion Framework \cite{ai_diffusion_framework}, which as one of its key directives proposed a host of requirements on models above $10^{26}$ FLOP (named ``controlled models'' in the Framework) with the aim of maintaining leadership in AI technology among the US and its allies \cite{heim2025ai}.

However, it is well established that the training compute used for frontier models has been growing extraordinarily quickly, with mean model size growing by about 4-5x per year over the past decade \cite{frontier_models_growth}. This has important implications for compute-based governance approaches such as those included in the EU AI Act and the Diffusion Framework. In April 2025, estimates suggest that there are 2 publicly available models trained using more than $10^{26}$ FLOP and approximately 30 publicly available models trained using more than $10^{25}$ FLOP \cite{epoch1e25update_tweet}. However, if current trends continue, these numbers may quickly grow. Governments will need to take this growth into account as they shape their AI governance efforts. By underestimating the number of models covered by their thresholds, they may fail to build sufficient capacity to implement their regulations or may impose regulatory burdens on an excessive number of actors.

With this in mind, we attempt to estimate the number of released models that will exceed various compute thresholds over the coming years. Extrapolating from current trends we conclude that by the end of 2028 there could be between \hyperref[results]{103-306 models} exceeding the $10^{25}$ FLOP threshold (90\% confidence interval) with a median estimate of 165, and \hyperref[results]{45-148 models} exceeding the $10^{26}$ FLOP threshold (90\% confidence interval), with a median estimate of 81. We also study ``frontier-connected thresholds'' -- thresholds that are defined relative to the largest training run at any one point in time rather than based on the absolute amount of training compute used -- and estimate that in the coming years there will be \hyperref[section:frontier-connected-results]{between 6-35 models released within 1 order of magnitude (OOM) of the largest training run that has taken place (90\% CI) with a stable median of 14-16 models} captured by this definition. However our analysis has limitations resulting from \hyperref[notable-models-selection-effect]{selection effects in the database that we extrapolate trends from}, as well as \hyperref[parameter-uncertainty-section]{uncertainty in key parameters that influence the projections}.

Importantly, our estimates do not straightforwardly translate into the number of models in scope of the EU AI Act or AI Diffusion Framework. Our numbers may provide an overestimate in that neither the EU nor the US would apply regulations to models trained and only made available in other jurisdictions (e.g., China). Additionally, the AI Act only applies to general purpose AI - it is unclear whether image and video generation models (such as OpenAI's SORA \cite{openai_sora} or Google DeepMind's Imagen \cite{google_deepmind_imagen}) would count as GPAI. The Act may also not apply to models that were placed on the market before the relevant obligations come into force in August 2025. Further, both rules could affect the market for AI development, making it less attractive for companies to release in-scope models. If these effects occur, our analysis may end up overcounting the models captured by these thresholds. At the same time, our estimates may underestimate the number of models subject to the compute-based threshold requirements. This is because the regulation could apply not only to original GPAISR developers but also to companies that modify the models \cite{williams2025regulating}, for example by adding software scaffolding around the model before making an application available to users. Since our analysis does not account for such adaptations, the actual number of affected models could be higher than our estimates suggest. Finally it should be noted that the EU AI Office has the ability to update the threshold in both directions - as does the US Bureau of Industry and Security in the U.S. as it relates to the AI Diffusion Framework - which would have to be taken into consideration when interpreting the predictions in future years.

\begin{figure}[H]
\centering
\fbox{\includegraphics[width=0.8\textwidth]{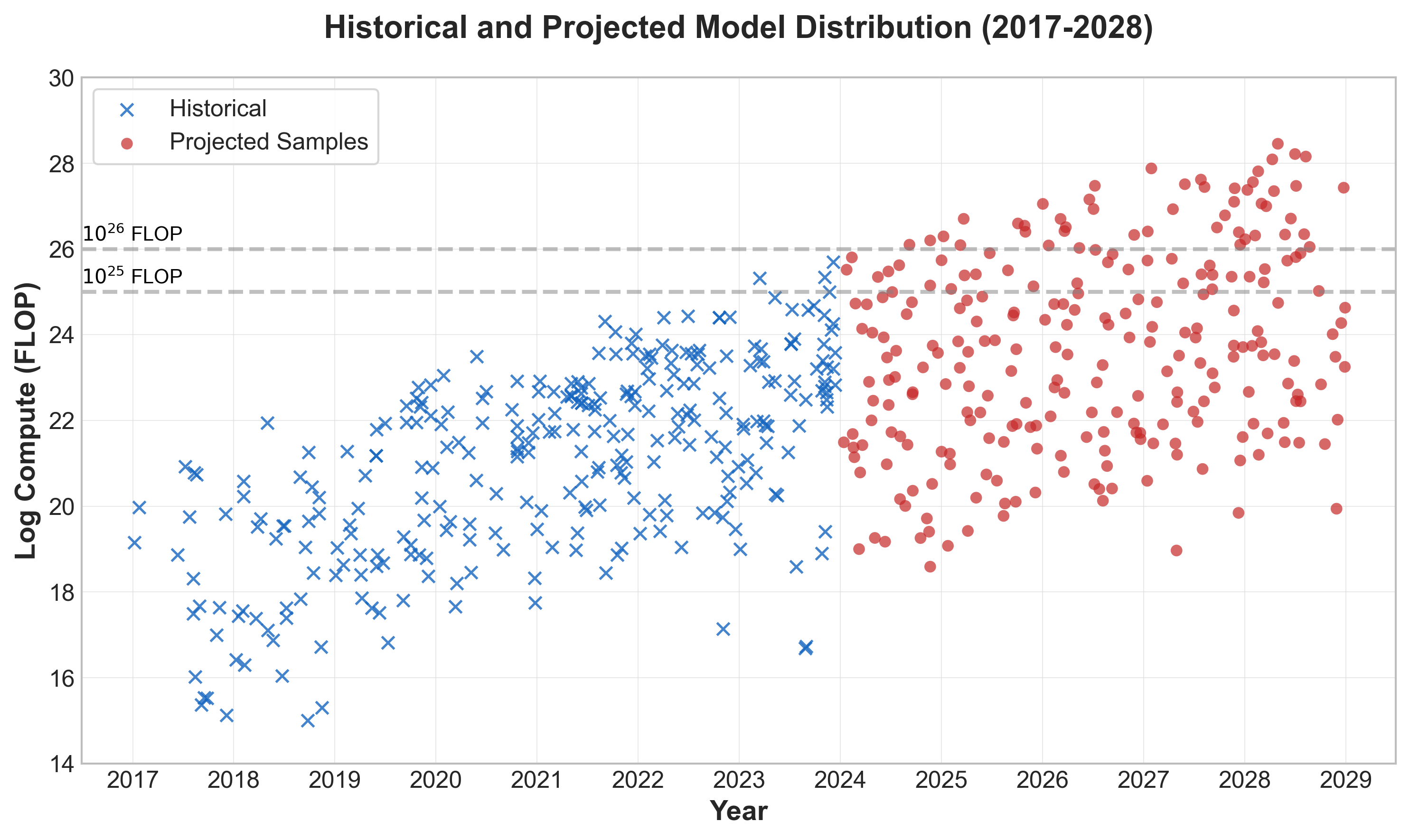}}
\caption{Historical data (2017-2023; blue) and a sample of our model’s predictions (2024-2028; red) for the number of AI models exceeding $10^{25}$ and $10^{26}$ FLOP. }
\label{fig:model-projections}
\end{figure}

%% file: texs/2_methodology.tex
\section{Methodology}

Our aim is to forecast the number of models that will be released above different training compute thresholds over the next four years. To do this, we model scenarios for the distribution of AI model releases over training compute. Once these distributions are established, we can simply count the models that exceed each specific compute threshold.

\subsection{Data}

We use Epoch AI's Notable Models dataset \cite{notable_ai_models} as the main dataset for our analysis. To our knowledge, this is the most comprehensive publicly available dataset of machine learning models available with over 450 entries of models and their estimated training compute. However, it is important to note that ML models must satisfy one of the following criteria (the notability criteria \cite{notable_models_inclusion} to be included in the database:

\begin{itemize}
\item highly cited (over 1000 citations);
\item large training cost (over \$1,000,000, measured in 2023 USD);
\item significant use (over one million monthly active users);
\item state of the art performance (typically on a recognized ML benchmark);
\item indisputable historical significance.
\end{itemize}

This makes the notable models dataset a strict subset of all machine learning models, and the selection effect applied by the notability criteria should be accounted for when interpreting the results in this paper. In short, this selection effect biases our median estimates towards being a lower bound estimate, especially for models that are multiple orders of magnitude of training compute below frontier models\footnote{This is because models multiple orders of magnitude away from the frontier are less likely to be captured by the notable model criteria relative to models close to the frontier.}. For these models that are significantly smaller than the frontier models, it is also difficult to quantitatively estimate the extent to which the estimates derived from the notable models database will be lower than the actual value. However, anchoring on the full 90\% prediction intervals presented in Section \ref{results} provides some insurance against the effects of this selection effect (i.e: in some cases the 95th percentile projection is likely to be closer to the actual number of models exceeding the compute threshold). The effects of the Notable Model database selection effects, and how this can be accounted for within our model are discussed in further detail in Section \ref{notable-models-selection-effect}.

Whilst the Notable Models database contains $\sim$450 entries with associated training compute values from 2010-2024, we fit our model based on the data from 2017-2023 (with 296 datapoints) to capture recent trends in the field. We do not use 2024 data to fit the model as the data is more likely to be incomplete - this is because the models tracked in the database lag behind models released by a handful of months. Visualizations of the model distribution in the dataset provide evidence in support of this, and are shown in appendix \ref{appendix-a-historical-distribution-of-notable-models-and-fit-data-choice}.

\subsection{Allocating total compute spending across different model scales}

We forecast the distribution of models over training compute by projecting the total amount of compute that will be spent training AI models in the coming years, and then modeling how this compute stock is distributed over models of different size. We choose this approach as we expect the total training compute spending to be a strong and relatively simple-to-forecast constraint on model distributions (over training compute). The allocation of compute to models of different sizes also exhibits a linear trend that is simple to extrapolate (discussed below).

Note that there are alternate ways to estimate the future distribution of models over training compute; for example, one could fit a parametric distribution (e.g: a normal distribution) to historical data, extrapolate this into future years and then sample from this distribution. However we do not use this method for the following reasons: Firstly, it is unclear which parametric distributions, if any, are a good fit to historical distributions of training compute (see appendix \ref{appendix-a-historical-distribution-of-notable-models-and-fit-data-choice}). Secondly, our approach has a more explicit focus on determining model distributions from total training compute spending - which, as mentioned above, we expect to be a strong and simple-to-forecast constraint over the next few years.

Using our approach, we project future model distributions over training compute by:

\begin{enumerate}
\item Projecting the total compute that will be used for AI workloads (both training and inference) with a median annual growth rate of roughly 4.1x.

\item Allocating this compute with a split of 40\% of compute towards training models that are publicly released and 60\% of compute towards other uses (including inference and compute used for research experiments) in 2025 and 2026, and a 30-70 allocation in 2027 and 2028.

\item Allocating the training compute across models of different scales - i.e: models within 1 OOM of the frontier model, models within 1 and 2 OOMs of the frontier model, models within 2 and 3 OOMs of the frontier model etc. - by fitting to data from 2017-2023 and assuming these allocation trends hold over the coming years.
\end{enumerate}

We will now discuss each of these in turn in greater detail.

We first begin by projecting the total amount of compute that will be used for AI workloads in the coming years. We use two sources for this.

Firstly, we can look at historical growth rates of training compute from the Notable Models database. Doing so for the years that we fit the model (2017-2023), we find a rapid growth rate of 6.3x annually for the total training compute used to train AI models. Assuming that the historical allocation between model training and inference has stayed roughly constant, we can generalise this to a 6.3x growth in compute used for AI workloads.

On the other hand, \cite{compute_forecast} models the growth in compute for AI workloads as increasing at a rate of 3.4x per year. This is the compound growth rate resulting from a 2.25x increase per year of the compute stock that can be used for AI workloads, and a 1.5x increase per year of the share of this stock that actually is allocated to AI training, inference and other uses\footnote{One can think of these two quantities as AI compute capacity and AI compute usage.}. The 2.25x growth rate in global AI-relevant compute stocks results from a 1.35x growth rate in the physical stock of AI chips and 1.65x in chip efficiency. The 1.5x growth in share results from aggressive buildouts of data centers by leading AI developers, financed by revenues on the order of tens of billions of dollars resulting from highly performant AI models and agents.

Given the discrepancy between these two estimates we integrate both into our median growth rate forecast. We put more weight on the 3.4x per year increase in compute for AI workloads given the detailed analysis that leads to this figure, however we do not fully discount the rapid growth rate that has been historically observed. Specifically we give the 3.4x figure three times as much weight as the historically observed figure, but this weighting is subjective and predictions for alternative weightings are shown in Appendix \ref{appendix-g-results-for-varying-growth-rate-weightings}. Applying this weighting between the two growth rates leads to a median growth rate of the AI workload stock (i.e: the stock of compute used for AI training and inference) of 4.1x per year. To account for uncertainty in the actual annual growth rate of compute for AI workloads, we add noise to the median growth rate, drawn from a normal distribution with mean of 0 and standard deviation of 0.5.

Next, we model (a) how the AI compute stock is allocated between training models, and other uses (such as model inference and research experiments) and (b) how the total training compute is allocated across models of different sizes. Our baseline scenario for part (a) follows the allocations in a recently released analysis (\cite{compute_forecast}), and is discussed in Section \ref{allocating-compute-between-training-inference-and-other-workloads}. To answer (b), we look at how training compute has been allocated to models of different sizes in recent years, and assume that these allocation trends hold in the coming years. This approach means that we do not have to explicitly commit to fixed parametric distributions.

Figure \ref{compute_allocations_2020_2023_unnormalised} below shows how training compute spending in the years 2020-2023 has been allocated to models of different sizes (data for 2017-2019 are shown in Appendix \ref{appendix-b-compute-allocations-for-2017-2019}). The x-axis represents the size of individual models. The y-axis shows the cumulative distribution function of training compute over model size - i.e: the fraction of training compute spent training models of size m or less. Differences in model sizes and their compute share can span orders of magnitude so both axes are log-scaled.

\begin{figure}[h]
\centering
\fbox{\includegraphics[width=0.8\textwidth]{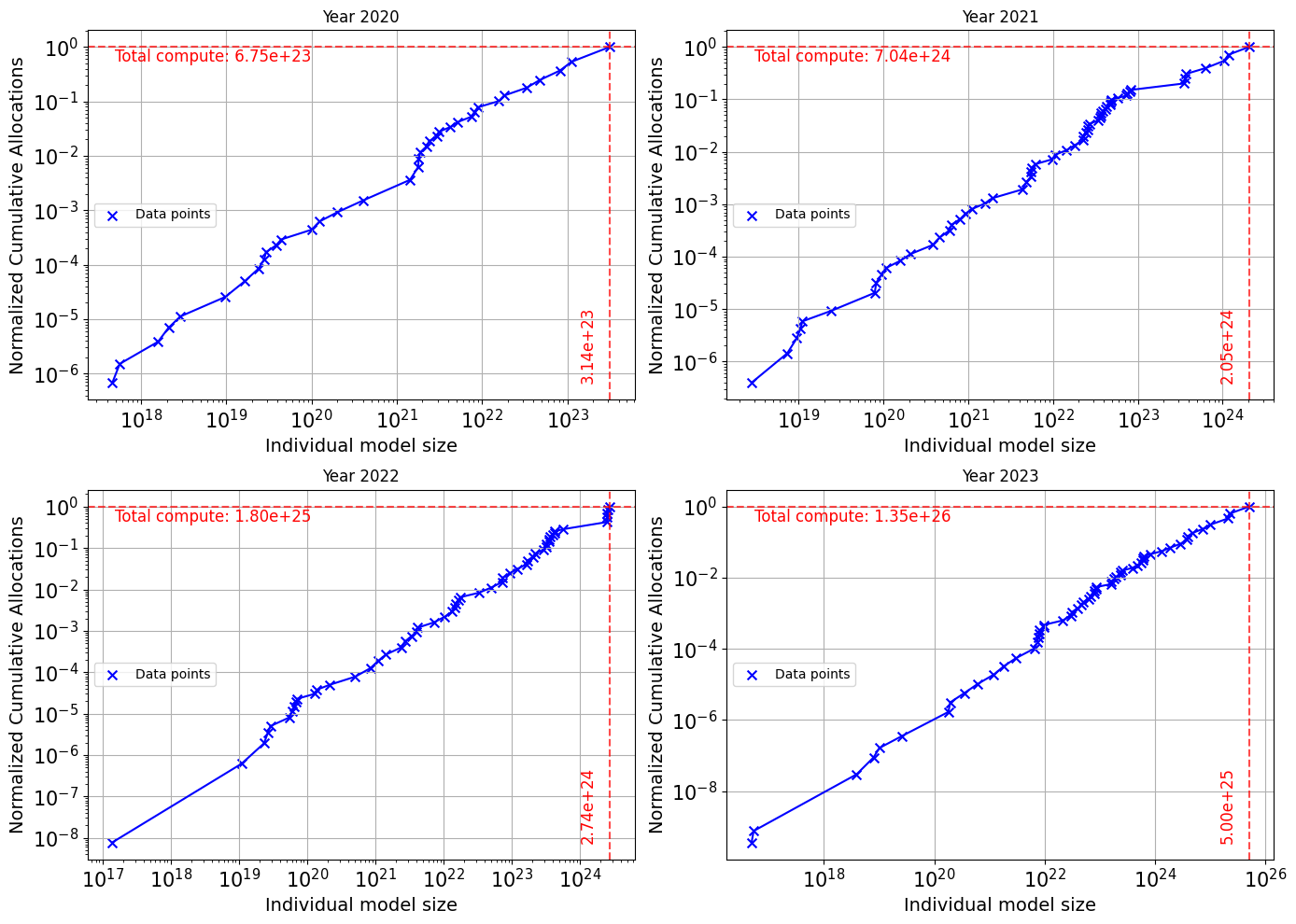}}
\caption{Compute allocation across model sizes for years 2020-2023, x-axis unnormalized.}
\label{compute_allocations_2020_2023_unnormalised}
\end{figure}

To get a concrete sense of what the plot shows, here are some conclusions that can be made by reading off the graph:

\begin{itemize}
\item In 2021, models of size $10^{22}$ FLOP or less contributed to approximately 1\% of total training compute spending.
\item In 2021, models of size $10^{21}$ FLOP or less contributed approximately 0.1\% of total training compute spending
\item In 2022, models of size $10^{21}$ FLOP or less contributed approximately 0.01\% of total training compute spending
\item In 2023, models of size $10^{23}$ FLOP or less contributed just less than 10\% total training compute spending.
\end{itemize}

Figure \ref{compute_allocations_2020_2023_unnormalised} also marks the largest model trained each year with the red vertical line. By definition, the largest model and all those smaller than it account for 100\% of training compute spending in a given year; this is shown in Figure \ref{compute_allocations_2020_2023_unnormalised} by the fact that this line intersects with the line representing total training compute spending.

\begin{figure}[h]
\centering
\fbox{\includegraphics[width=0.8\textwidth]{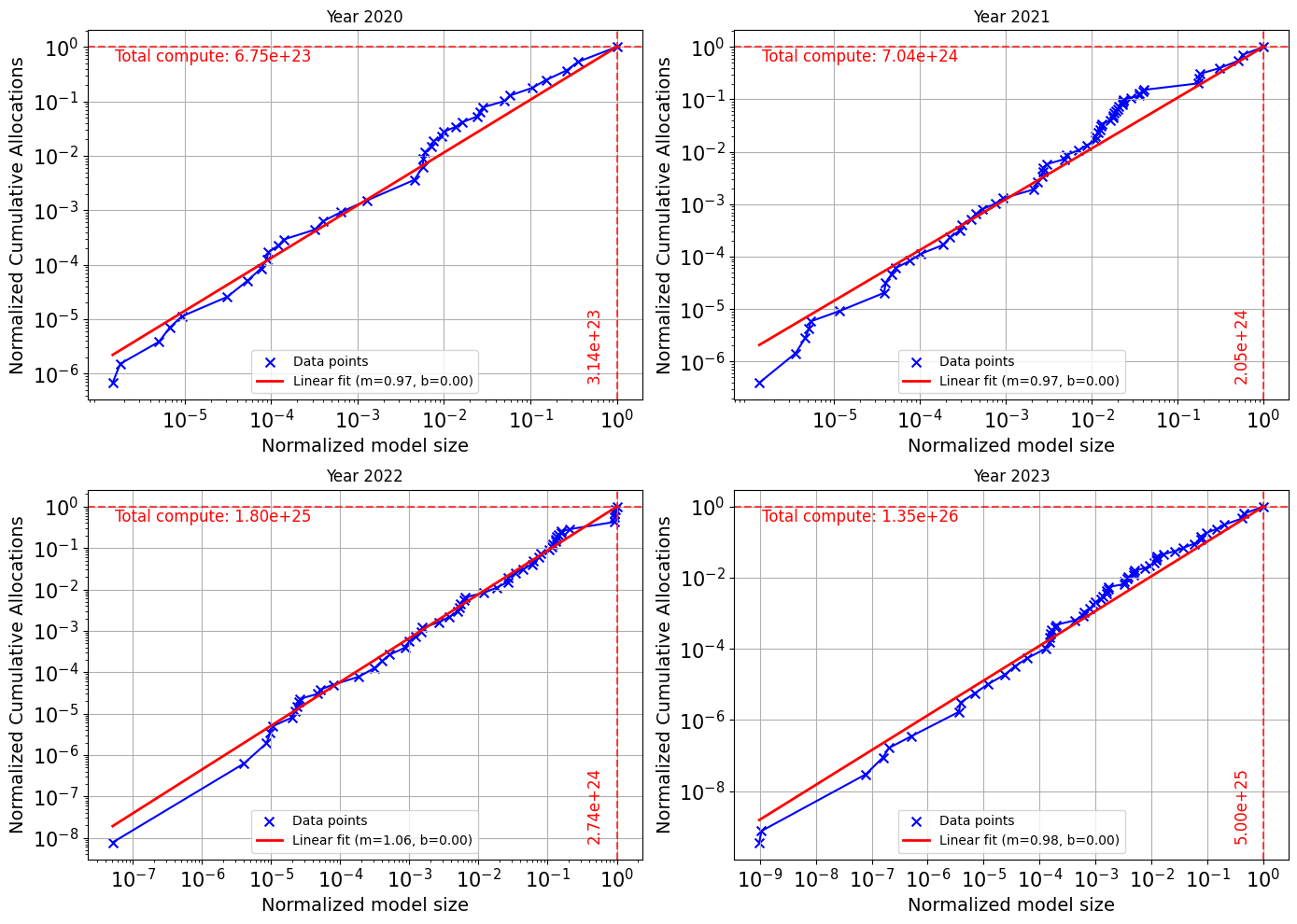}}
\caption{Compute allocation among model sizes for years 2020-2023, x-axis normalized by largest model trained that year. Linear fit shown.}
\label{figure:compute_allocations_2020_2023_normalised}
\end{figure}

The relationship between model size m and the cumulative distribution function is consistently linear across 2020-2023 (and 2017-2019 - see Appendix \ref{appendix-b-compute-allocations-for-2017-2019}), suggesting a stable trend that can be extrapolated. The size of individual models grows each year, so to extrapolate this trend we normalize the x-axis by the largest model trained in each year (shown in Figure \ref{figure:compute_allocations_2020_2023_normalised}). Table \ref{tab:2023-allocations} shows the compute allocations for different model sizes for 2023 that are derived from these plots.

\begin{table}[H]
\centering
\label{table-1-allocations}
\small
\setlength{\tabcolsep}{5pt}
\renewcommand{\arraystretch}{1.3}
\begin{tabular}{>{\raggedright\arraybackslash}p{2.2cm}|>{\centering\arraybackslash}p{2cm}|>{\centering\arraybackslash}p{2cm}|>{\centering\arraybackslash}p{2cm}|>{\centering\arraybackslash}p{2cm}|>{\centering\arraybackslash}p{2cm}}
\toprule
\multicolumn{6}{c}{\textbf{Compute allocation in 2023}} \\
\midrule
\textbf{Model size relative to Gemini Ultra} & 
All models within 5--4 OOM & 
All models within 4--3 OOM & 
All models within 3--2 OOM & 
All models within 2--1 OOM & 
All models within 1 OOM \\
\midrule
\textbf{Model size (absolute)} & 
5$\times$10$^{20}$--5$\times$10$^{21}$ & 
5$\times$10$^{21}$--5$\times$10$^{22}$ & 
5$\times$10$^{22}$--5$\times$10$^{23}$ & 
5$\times$10$^{23}$--5$\times$10$^{24}$ & 
5$\times$10$^{24}$--5$\times$10$^{25}$ \\
\midrule
\textbf{Fractional allocation} (2 s.f.) & 
0.00011\% & 
0.0010\% & 
0.010\% & 
8.6\% & 
90\% \\
\midrule
\textbf{Compute allocation} (FLOP) & 
1.51$\times$10$^{22}$ & 
1.43$\times$10$^{23}$ & 
1.36$\times$10$^{24}$ & 
1.16$\times$10$^{25}$ & 
1.22$\times$10$^{26}$ \\
\bottomrule
\end{tabular}
\vspace{0.3cm}
\caption{2023 allocations -- Largest model: Gemini Ultra @ 5×10$^{25}$ FLOP, total compute = 1.35×10$^{26}$ FLOP. OOM refers to an order of magnitude of training compute. Allocations may not sum exactly to 1 due to rounding errors.}
\begin{minipage}{\textwidth}
\end{minipage}
\label{tab:2023-allocations}
\end{table}

Before moving on, it is worth briefly considering the constraints and physical interpretations of the parameters of the linear fits - namely, the gradient (k) and the intercept (b). First, note that the linear fits must pass through (1,1) on the normalized plots (Figure 2B) - this is because, by definition, the largest model and all those smaller than it (i.e: all models released in a given year) account for 100\% of compute used. This means that the intercept of the linear fits - b - must be 0 (see Appendix \ref{appendix-c-constraints-and-interpretation-of-the-linear-fits} for details). Regarding k, Appendix \ref{appendix-c-constraints-and-interpretation-of-the-linear-fits} shows that models that are 10x the size of a smaller counterpart are allocated roughly 10$^k$ times as much compute. Historical values for k are seemingly equally distributed across the range [0.9, 1.1] (shown in Appendix \ref{appendix-d-historical-values-of-allocation-gradients}), therefore our model samples k uniformly from the range [0.9,1.1]. The edge case of k=0.9 corresponds to models that are 10x as large being allocated 10$^{0.9}$ = $\sim$8.0 times as much compute, whereas for the k=1.1 case this factor is 10$^{1.1}$ = 12.6 times as much compute. Table \ref{tab:compute-allocations-k-values} shows allocations of compute across model size for a range of different k. 

\subsection{Modelling the largest training run}
\label{section:lms_modelling_assumptions}

Recall from Table \ref{tab:2023-allocations} that our method assigns training compute to models based on their size relative to the largest training run. Therefore, assumptions must be made about the size of the largest training run in a given year, which is done by making assumptions about the share of total training compute that the largest training run uses. In our model, this parameter is called the largest model share (or LMS). To illustrate what the the LMS represents, setting an LMS of 0.3 (or 30\%) means that the largest model trained in a given year is uses 30\% of the total training compute spent used across all models that year. 

Given this setup, what are reasonable assumptions to make about the largest model share (LMS) parameter? One source of evidence is the historical values of the LMS derived from the Notable Models database; these are shown in Appendix \ref{f.1-historical-lms}. One can see a range of values - for example, in 2018 ResNet-101 used approximately 78\% of the training compute that was spent across all models tracked in the Notable Models database that year. More recently, in 2020 GPT-3 175B (davinci) accounted for ~46\% of compute spent across all models trained and released that year. On the lower end, data from 2022 shows Minerva 540B accounting for 15\% of training compute spent that year on released models. Discounting the 2018 LMS as an outlier, the historical values appear relatively evenly distributed over the range [0.1,0.5] - therefore sampling the LMS parameter uniformly from this range (LMS $\sim \text{Unif}(0.10,0.50) $) is a reasonable first modelling attempt. However, lower values of the LMS exert a strong influence on the model's predictions for reasons discussed in Appendix \ref{f.2-influence-of-lms-on-projections}, so the lower bound of the sampling interval is extended to 0.05 to yield a refined model of LMS $\sim \text{Unif}(0.05,0.50)$. 

The LMS parameter is sampled uniformly from the bounds [0.05, 0.50] when the model is retrodicted for the years 2020-2023 in Section \ref{section:verification}, however one more refinement is made when projecting the model forward for the years 2025-2028. A qualitative interpretation of the LMS parameter is the degree of concentration in the market of developers producing AI models at the largest scales. To illustrate, consider an LMS of 0.5, meaning that the largest training run used 50\% of the training compute spent across all models that year. An implication of this is that at most, two developers can train a model at this scale. On the other hand an LMS of 0.05 (i.e: the largest training run uses 5\% of training compute that was spent across all models released that year) means that, in theory, 20 different actors could participate in model development at this scale\footnote{The assumption here is that no actor releases multiple models at this scale, but the general argument still holds without this.}. 
We expect an increasing number of developers to perform training runs at the largest scales in the coming years. This trend can be seen today with a number of relative newcomers, such as x.AI, Inflection, and Mistral, joining established actors such as OpenAI and Google DeepMind in training models at the very frontier of AI development\cite{models_over_1e25}. To quantitatively model this trend of a greater number of actors participating in frontier AI development, we sample the LMS parameter log-normally for the years 2025-2028\footnote{Formally, $\log{(LMS)} \sim \mathcal{N}(\mu,\sigma)$ where $\mu = \mathbb{E}(\log(0.05),\log(0.50))$, $\sigma = (\log(0.50)-\log(0.05))/4$} from the bounds discussed above, in contrast to the uniform sampling used when retrodicting the model for 2020-2023 (these two LMS sampling distributions can be seen in Appendix \ref{appendix-results-uniform-sampling}, alongside results using uniform sampling of the LMS for the years 2025-2028). For 2024, the largest model released in 2024 is documented in the Notable Models database as GPT-4o \cite{hurst2024gpt} at $3.8 \times 10^{25}$ FLOP, hence we set the LMS for 2024 to match this. 

Finally, to generate model distributions to extend the historical data we randomly sample models from each bin until the total training compute allocated to that bin is met or exceeded.

\subsection{Allocating compute between training, inference, and other workloads}
\label{allocating-compute-between-training-inference-and-other-workloads}

Given estimates of the total compute stock, the next key stage involves allocating this stock between training, inference and other workloads. First a note on the different usages for compute. AI developers can use compute for model training (both pre- and post-training), as well as serving models to customers (external deployment/inference), using models in-house for research automation and monitoring (internal deployment/inference), compute for research experiments, generating synthetic data, and more. This analysis is primarily concerned with the compute used for training released models, hence compute usage is categories into 'training' and 'inference and other'.

\cite{compute_forecast} estimates that in 2024 approximately 40\% of compute was used for training (including both pre- and post-training), with the remaining share going to model inference and other uses. This is used as the starting point for the model. \cite{compute_forecast} then estimate the share of compute allocated to training in the following years, estimating that in 2025 and 2026 40\% of compute will be used for model training, 30\% at the start of 2027 and 20\% by the end of 2027. These allocations are mapped to the years of predictions for this forecast (2025-2028) in the first column of Table \ref{compute_allocations_baseline}\footnote{Concretely, we take the start of 2027 allocation (30\%) as the allocation for all of 2027, and the end of 2027 allocation (20\%) as the 2028 allocation.}. The 20-80 allocation of compute between training and other uses at the end of 2027 represents an aggressive scenario with respect to AI automation, and is therefore adjusted to a slightly more balanced 30-70 split for the baseline forecast in this paper, as can be observed in the second column of Table \ref{compute_allocations_baseline}.

It is also worth noting that there is considerable uncertainty in the estimated present day compute allocations between training, inference and other workloads, with public estimates varying widely\footnote{For example, the Epoch GATE model \cite{gate_model} estimates an allocation of 90\% of compute to model training in 2024, moving to 70\% in 2028.}. In Appendix \ref{appendix-h---results-for-alternate-training-compute-allocations} we present results for alternate allocation scenarios.

\begin{table}[htbp]
\centering

\renewcommand{\arraystretch}{1.3}
\begin{tabular}{>{\raggedright\arraybackslash}p{1.5cm}>{\centering\arraybackslash}p{5cm}>{\centering\arraybackslash}p{5cm}}
\toprule
\textbf{Year} & \textbf{Approximate training compute allocations} \cite{compute_forecast} & \textbf{Training compute allocation} (ours) \\
\midrule
2025 & 40\% & 40\% \\
2026 & 40\% & 40\% \\
2027 & 30\% & 30\% \\
2028 & 20\% & 30\% \\
\bottomrule
\end{tabular}
\caption{Compute allocations for model training (not including research experiments) for 2025-2028}
\label{compute_allocations_baseline}
\end{table}

\subsection{Frontier-connected thresholds}
\label{frontier-connected-thresholds}

An alternative to setting training compute thresholds based on absolute compute limits (e.g: 10$^{25}$ FLOP, 10$^{26}$ FLOP) is to set thresholds based upon a model's proximity to the largest model trained to date\footnote{See Appendix A of 
 \cite{anderljung2023frontier} for discussion.}. To get a sense for how such a threshold would operate, at a given point in time a model regulator could require that any model trained over the next 3 months that is either (a) within 1 OOM of the largest model that existed at the start of this period or (b) exceeding the size of this largest model is subjected to additional reporting or transparency requirements. We project trends for these ‘frontier-connected’ thresholds for models that are within 0.5, 1 and 1.5 orders of magnitude of the largest model to date.

%% file: texs/3_results.tex
\section{Results}
\label{results}

In this section we present the results of our model for both absolute compute thresholds (e.g: 10$^{25}$ FLOP, etc.) and for frontier-connected thresholds that incorporate models within 0.5, 1.0 and 1.5 orders of magnitude from the largest model released. We present the results of our model in the format (5, 50, 95), where 5, 50, 95 refer to the 5th, 50th and 95th percentile projections of the model when running it 1000 times. The 90\% confidence prediction intervals are presented to convey uncertainty about parameters of the model, such as the largest model share (LMS), compute growth rates and allocation gradient. Interpretations of these results should anchor on the full prediction interval, rather than a single point estimate (e.g: the median projection).

Section \ref{section:verification} aims to validate the model projections by retrodicting our model and comparing the results to the historically observed data. Here we find that the 90\% prediction intervals capture all historically observed data points. Taken together, this provides considerable evidence that the 90\% prediction intervals presented below are likely to capture the number of models released in the coming years above each respective threshold.\footnote{Supporting code for this analysis will be made available at \url{https://github.com/IyngkarranKumar/compute_thresholds_public}.
}

\subsection{Absolute compute thresholds}

First, we can compare our model's projections to existing data on the number of AI models exceeding 10$^{25}$ FLOP. The model predicts that there were 23 AI models exceeding 10$^{25}$ FLOP at the end of 2024, which aligns well with the 24 found in \cite{models_over_1e25} (recall that we set the largest model in 2024 to the size of GPT-4o inline with the Notable Models database). Then, assuming these prediction intervals will hold over the coming years, what are the implications for training compute thresholds? By the end of 2028 (EOY 2028) our median and 95th percentile projections for the number of models exceeding the 10$^{25}$ FLOP threshold in the EU AI Act are 165 and 306 respectively. Our median and 95th percentile projection for the number of models exceeding the 10$^{26}$ threshold in the Diffusion Framework are 81 and 148.

\begin{table}[htbp]
\centering
\renewcommand{\arraystretch}{1.3}
\begin{tabular}{>{\raggedright\arraybackslash}p{2.2cm}>{\centering\arraybackslash}p{2.2cm}>{\centering\arraybackslash}p{2.2cm}>{\centering\arraybackslash}p{2.2cm}>{\centering\arraybackslash}p{2.2cm}>{\centering\arraybackslash}p{2.2cm}}
\toprule
\textbf{Threshold (FLOP)} & \textbf{2024} & \textbf{2025} & \textbf{2026} & \textbf{2027} & \textbf{2028} \\
\midrule
$\boldsymbol{>}$ \textbf{10$^{25}$} & \mbox{[18, 23, 27]} & \mbox{[32, 45, 64]} & \mbox{[51, 77, 119]} & \mbox{[76, 117, 201]} & \mbox{[103, 165, 306]} \\
$\boldsymbol{>}$ \textbf{10$^{26}$} & \mbox{[0, 0, 0]} & \mbox{[3, 7, 11]} & \mbox{[12, 24, 38]} & \mbox{[27, 47, 81]} & \mbox{[45, 81, 148]} \\
$\boldsymbol{>}$ \textbf{10$^{27}$} & \mbox{[0, 0, 0]} & \mbox{[0, 0, 0]} & \mbox{[0, 2, 5]} & \mbox{[1, 10, 20]} & \mbox{[9, 27, 56]} \\
$\boldsymbol{>}$ \textbf{10$^{28}$} & \mbox{[0, 0, 0]} & \mbox{[0, 0, 0]} & \mbox{[0, 0, 0]} & \mbox{[0, 0, 0]} & \mbox{[0, 3, 8]} \\
$\boldsymbol{>}$ \textbf{10$^{29}$} & \mbox{[0, 0, 0]} & \mbox{[0, 0, 0]} & \mbox{[0, 0, 0]} & \mbox{[0, 0, 0]} & \mbox{[0, 0, 0]} \\
\bottomrule
\end{tabular}
\caption{Results for absolute thresholds. The table presents 90\% prediction intervals [5th, 50th, 95th percentile] for the number of models exceeding each compute threshold. Results are cumulative, showing estimates for the number of models released by the end of each year.}
\label{tab:absolute-thresholds-baseline}
\end{table}

It is also worth noting the growth rate of the number of models exceeding the AI Act's threshold each year. Considering the 95th percentile projections, we see that from 2024-2025 the number of models captured by the 10$^{25}$ threshold increases by 37, and then 55, 82 and 105 in subsequent years. This highlights a superlinear growth trend; not only does the number of models captured by the Act's compute threshold increase, but it increases at a growing rate. This superlinear growth trend holds across all projections (5th, 50th and 95th percentile forecasts) and across all absolute compute thresholds. However, the multiplicative factor from one year to the next is decreasing, meaning the growth is sub-exponential. For example, the multiplicative factor between median estimates for the 10$^{25}$ FLOP threshold is 1.96, 1.71, 1.52 and 1.41. Once again, this holds across all thresholds and percentile estimates.

\subsection{Frontier-connected thresholds}
\label{section:frontier-connected-results}

We now consider the results of our projections for the frontier-connected thresholds defined in \ref{frontier-connected-thresholds}. These results are shown in Table \ref{tab:frontier-thresholds-baseline} The key takeaway is that across all projections (5th, 50th and 95th percentile) and all thresholds (within 0.5, 1.0 and 1.5 OOMs of the largest model), the number of models captured by the threshold from 2025-2028 stays roughly constant. The figures differ for 2024 as the largest training run is fixed at GPT-4o's size of $3.8\times10^{25}$ FLOP. For example, our median projection for the number of models that are within 1 OOM of the largest model at a given date is stable at 14-16 models across each year from 2025-2028, and the other thresholds of 0.5 OOMs and 1.5 OOMs see a similar trend, with stable medians in the range of 6-8 and 20-24 respectively. This constancy in the number of models captured by the threshold may be a desirable property for regulatory bodies responsible for enforcement; however the extent to which the frontier-connected thresholds are preferable over absolute training compute thresholds will also depend on whether the risks posed by a model are in part due to the absolute amount of compute used to train the system or the amount used relative to the frontier.

\begin{table}[H]
\centering
\renewcommand{\arraystretch}{1.3}
\begin{tabular}{>{\raggedright\arraybackslash}p{2.5cm}>{\centering\arraybackslash}p{2cm}>{\centering\arraybackslash}p{2cm}>{\centering\arraybackslash}p{2cm}>{\centering\arraybackslash}p{2cm}>{\centering\arraybackslash}p{2cm}}
\toprule
\textbf{Distance from frontier model} & \textbf{2024} & \textbf{2025} & \textbf{2026} & \textbf{2027} & \textbf{2028} \\
\midrule
\textbf{Within 0.5 OOM} & \mbox{[11, 16, 21]} & \mbox{[3, 6, 14]} & \mbox{[2, 7, 16]} & \mbox{[3, 8, 15]} & \mbox{[3, 8, 17]} \\
\textbf{Within 1 OOM} & \mbox{[23, 31, 39]} & \mbox{[7, 14, 25]} & \mbox{[6, 15, 29]} & \mbox{[7, 16, 31]} & \mbox{[7, 15, 35]} \\
\textbf{Within 1.5 OOM} & \mbox{[36, 46, 60]} & \mbox{[12, 20, 39]} & \mbox{[11, 22, 46]} & \mbox{[11, 24, 50]} & \mbox{[11, 23, 55]} \\
\bottomrule
\end{tabular}
\caption{Frontier-connected thresholds. Results show 90\% prediction intervals [5th, 50th, 95th percentile] for models within specified distances of the frontier model. Results for each year represent new models released in that year only.}
\label{tab:frontier-thresholds-baseline}
\end{table}

%% file: texs/4_verification.tex
\section{Verification}
\label{section:verification}

In this section we present results obtained by retrodicting our model for absolute compute thresholds 10$^{23}$, 10$^{24}$ and 10$^{25}$ FLOP, for which there exists data from the years 2020-2023. We also compare our models predictions with the number of AI models satisfying the frontier-connected thresholds. The model's 90\% prediction intervals capture all of these historically observed data points, which provides considerable evidence that the predicted intervals in Section \ref{results} will capture the actual number of models exceeding the thresholds.

\subsection{Absolute compute thresholds}

\begin{table}[H]
\centering
\renewcommand{\arraystretch}{1.3}
\begin{tabular}{>{\raggedright\arraybackslash}p{2.2cm}>{\centering\arraybackslash}p{2.2cm}>{\centering\arraybackslash}p{2.2cm}>{\centering\arraybackslash}p{2.2cm}>{\centering\arraybackslash}p{2.2cm}}
\toprule
\textbf{Threshold (FLOP)} & \textbf{2020} & \textbf{2021} & \textbf{2022} & \textbf{2023} \\
\midrule
$\boldsymbol{>}$ \textbf{10$^{23}$} & \mbox{2 (0,1,4)} & \mbox{9 (8,13,27)} & \mbox{29 (19,29,60)} & \mbox{54 (36,54,128)} \\
$\boldsymbol{>}$ \textbf{10$^{24}$} & \mbox{0 (0,0,0)} & \mbox{3 (0,2,3)} & \mbox{8 (3,8,10)} & \mbox{19 (13,22,44)} \\
$\boldsymbol{>}$ \textbf{10$^{25}$} & \mbox{0 (0,0,0)} & \mbox{0 (0,0,0)} & \mbox{0 (0,0,0)} & \mbox{4 (0,4,5)} \\
\bottomrule
\end{tabular}
\caption{Absolute compute thresholds retrodiction. Each cell is formatted as $\text{O } (5,50,95)$ where $\text{O},5,50,95$ are the historically observed values, 5th percentile, 50th percentile (median) and 95th percentile prediction.}
\label{tab:absolute-retrodiction}
\end{table}

For the absolute compute thresholds 10$^{23}$, 10$^{24}$ and 10$^{25}$ FLOP, our 90\% confidence interval captures the historically observed values for the years 2020-2023. Note however, that for a number of cells the historically observed value is closer to the model's 5th or 95th projection than it is to the median projection, highlighting the importance of interpreting the full prediction interval.

\subsection{Frontier-connected threshold}

\begin{table}[hbtp]
\centering
\renewcommand{\arraystretch}{1.3}
\begin{tabular}{>{\raggedright\arraybackslash}p{2.5cm}>{\centering\arraybackslash}p{2.2cm}>{\centering\arraybackslash}p{2.2cm}>{\centering\arraybackslash}p{2.2cm}>{\centering\arraybackslash}p{2.2cm}}
\toprule
\textbf{Distance from frontier model} & \textbf{2020} & \textbf{2021} & \textbf{2022} & \textbf{2023} \\
\midrule
\textbf{Within 0.5 OOM} & \mbox{3 (1,6,12)} & \mbox{4 (1,4,19)} & \mbox{5 (1,5,12)} & \mbox{5 (2,4,18)} \\
\textbf{Within 1.0 OOM} & \mbox{7 (5,12,24)} & \mbox{19 (5,9,34)} & \mbox{16 (4,9,28)} & \mbox{11 (5,9,35)} \\
\textbf{Within 1.5 OOM} & \mbox{11 (8,18,39)} & \mbox{27 (8,15,51)} & \mbox{22 (7,15,44)} & \mbox{17 (9,14,50)} \\
\bottomrule
\end{tabular}
\caption{Frontier-connected thresholds retrodiction. Each cell is formatted as $\text{O} (5,50,95)$ where $\text{O},5,50,95$ are the historically observed values, 5th percentile, 50th percentile (median) and 95th percentile prediction.}
\label{tab:frontier-retrodiction}
\end{table}

For the 0.5, 1.0 and 1.5 OOM frontier-connected thresholds, our 90\% confidence interval captures all of the historically observed values. Again, in some cases (e.g: 2021), the observed value is closer to the 5th or 95th percentile projection than it is to the median projection. 

%% file: texs/5_limitations.tex
\section{Limitations}

In this section the limitations of our analysis are discussed. This includes the selection effect on the machine learning models tracked by the Notable Models database, as well as limitations resulting from uncertainty in key parameters that influence the model's predictions (such as the largest model share parameter) and compute growth rates.

\subsection{Notable models selection effect}
\label{notable-models-selection-effect}

The Notable Models database is a strict subset of all machine learning models. Therefore using this dataset to model future trends will, in general, lead to point estimates of the number of models that exceed the absolute compute thresholds being a lower bound on the true number\textit{.} However, the nearer a model is to the frontier, the closer this lower bound estimate will be to the actual number of models that exceed the threshold. To illustrate, consider the cases of setting a 10$^{22}$ FLOP compute threshold and 10$^{25}$ FLOP threshold in the year 2023. In 2023, the largest model released was Gemini Ultra at 5x10$^{25}$ FLOP. Most models with training compute on the order of 10$^{25}$ are likely to have satisfied at least one of the notability criteria\footnote{As a reminder, the notability criteria are:
\begin{itemize}
\item highly cited (over 1000 citations);
\item large training cost (over \$1,000,000, measured in 2023 USD);
\item significant use (over one million monthly active users);
\item state of the art performance (typically on a recognized ML benchmark);
\item indisputable historical significance.
\end{itemize}} given their proximity to the frontier AI model of the day. However, models of size $\sim$10$^{22}$ FLOP in 2023 are far less likely to have met the notability criteria and thus more likely to be excluded from the dataset, making point estimates of the trends of 10$^{22}$ FLOP models less reliable.

One way in which this effect can be accounted for within our model is by setting the allocation gradient (k) to less than one. Qualitatively this means that smaller models get more compute relative to the median scenario in our baseline case (where the median value of k is 1.0). For example, in the case where k = 1.0, models that are within 1 OOM of the largest model trained in a given year are allocated 90\% of the compute share, with models within 1 OOM and 2 OOM of the largest model getting a factor of 10 less (i.e., 9\%). On the other hand, when k = 0.5, models within 1 OOM of the frontier get only 68\% of compute, and those in the category below it get 10$^{0.5}$ $\approx$ 3.2x less compute (other allocations are shown in Table \ref{tab:compute-allocations-k-values}). The results of our baseline model shown in Section \ref{results} choose k based on historical values found from the Notable Models database, as it is unclear how exactly to modify this parameter to account for the notable models selection effect. However in Appendix \ref{section:appendix-varying-allocation-gradient}, results are presented in which k is sampled uniformly from the ranges [0.7,0.9] and [0.5, 0.7] to illustrate the outcome on compute thresholds counts when using this technique as a potential correction for the notable models selection effect.

Even within our baseline predictions (where k is sampled uniformly from [0.9,1.1]), it should be emphasized that our 90\% prediction intervals are designed to capture uncertainty in the model parameters (such as largest model share, allocation gradient, and compute stock growth rate), not the uncertainty that arises from the Notable Models selection effect. However anchoring on the full prediction intervals will insure decisions against this selection effect to some extent. In other words, while our median projections are biased towards underestimating the actual number of models captured by the compute thresholds, this is less likely to be the case for our 95th percentile projections.

\subsection{Uncertainty in key model parameters}
\label{parameter-uncertainty-section}

Key parameters that influence the model's predictions are the largest model share parameter (LMS), the compute stock growth rate, the allocation gradient, and the allocation of the compute stock between training and inference. Our choices for most of these parameters are informed by their historical values, for which we have limited data (2017-2023). This introduces uncertainty into our model, which we account for by presenting 90\% confidence intervals. However the limited historical data often only provides six (6) data points to calibrate these intervals with.

For example, the LMS in previous years has varied significantly over the range 0.1 (2024) to 0.46 (2020). In making projections for the years 2025-2028 we therefore sample the LMS (lognormally) from a wider range of [0.05,0.5]. Given that low values of the LMS can lead to substantially different projections compared to larger values (Appendix \ref{f.2-influence-of-lms-on-projections} discusses this in further detail), future work could look to further calibrate this parameter as more data becomes available. The case for the compute stock growth rate is similar - the two sources of evidence that are used in this study (historical compute growth rates and \cite{compute_forecast}'s forecast of future compute stock growth) show a noticeable discrepancy which we resolve by subjectively weighting the estimated growth rates. Valuable future work could look to conduct further analysis into the growth of compute that is available for AI training and inference.

%% file: texs/6_discussion.tex
\section{Implications for Compute-Threshold-Based Governance Frameworks}

Training compute is increasingly being used as a proxy for determining which AI models should be subject to requirements imposed by AI governance frameworks. As \cite{heim_koessler_2024} argue, compute thresholds serve as an effective initial filter to identify potentially risky general-purpose AI models that warrant regulatory oversight and further scrutiny. This approach is valuable because training compute correlates with model capabilities and potential risks while being easily quantifiable compared to other inputs to AI development \cite{sastry_2024}.

However, our analysis demonstrates a critical challenge: any static compute threshold will capture an increasing number of models over time. By the end of 2028, our median projections indicate that 165 models will exceed the 10$^{25}$ FLOP threshold established in the EU AI Act, and 81 models will surpass the 10$^{26}$ FLOP threshold set in the US AI Diffusion Framework. Especially relevant is the superlinear growth pattern we observe. Not only does the number of models captured by these thresholds increase yearly, but the rate of increase accelerates.

This pace of increase could strain governmental capacity to enforce and monitor requirements while potentially imposing excessive regulatory burdens on a growing number of AI developers. In response, governments have at least two potential options. First, they can make regulatory burdens more proportionate for less risky models while increasing their regulatory capacity to handle the growing number of models in scope. Second, they can reduce the scope of requirements by excluding certain models from the regime. Both the EU AI Office in the EU and the Bureau of Industry and Security in the US have the authority to make such adjustments to the scope of their respective regulatory frameworks.

%% file: texs/7_conclusion.tex
\section{Conclusion}

In this paper we estimate the number of AI models that will exceed training compute thresholds such as those proposed in the EU AI Act \cite{eu_ai_act} and the AI Diffusion Framework \cite{ai_diffusion_framework}. Our method centres around estimating the total stock of compute that will be available for AI workloads (training and inference), and then allocating the training compute stock to models of different size following trends seen from 2017-2023. We find that 103-306 models will exist by the end of 2028 that exceed a 10$^{25}$ FLOP compute threshold with a 90\% confidence interval, and 45-148 models will exist above the 10$^{26}$ FLOP threshold. For all compute thresholds defined with respect to an absolute value, the number of models exceeding the threshold increases at an accelerating (superlinear) rate.

We also analyse trends for compute thresholds that are connected to the largest frontier model at the time of their release and find for thresholds that capture models within 0.5, 1.0 and 1.5 orders of magnitude of the largest model, the median number of models captured remains roughly stable each year (at approximately 7, 15 and 21-22 respectively) - however we have wide 90\% confidence intervals around these. We validate our predictions by retrodicting threshold counts for the years 2020-2023, with our 90\% confidence intervals capturing all historically observed values.

However our analysis is limited by the selection effects applied by the Notable Models database which bias our median projections towards being a lower bound on the actual number of models that will exceed the absolute compute thresholds. Additionally uncertainty around model parameters such as the largest model share (LMS), allocation gradient and compute growth rate mean that the full 90\% prediction interval should be accounted for when basing policy decisions on these results. These findings emphasize the need for policymakers and regulatory bodies to consider the rapid growth in frontier AI model counts when designing and implementing compute-based governance frameworks, ensuring they have sufficient capacity to monitor and regulate an expanding number of models while maintaining flexibility to adjust thresholds as the AI landscape evolves.

%% file: texs/11_acknowledgments.tex
\section*{Acknowledgements}
For thoughtful feedback and discussions related to this work, we would like to thank: Markus Anderljung, David Owen, Ben Cottier, Josh You, Lennart Heim, Rani Martin, John Halstead, Alan Chan, Oscar Delaney, and the entire 2025 Winter Fellow cohort at GovAI.

%% file: texs/8_appendix_a_c.tex
\newpage
\appendix
\section*{Appendices}
\addcontentsline{toc}{section}{Appendices}

\section{Historical distribution of notable models and fit data choice}
\label{appendix-a-historical-distribution-of-notable-models-and-fit-data-choice}

\begin{figure}[h]
\centering
\fbox{\includegraphics[width=0.6\textwidth]{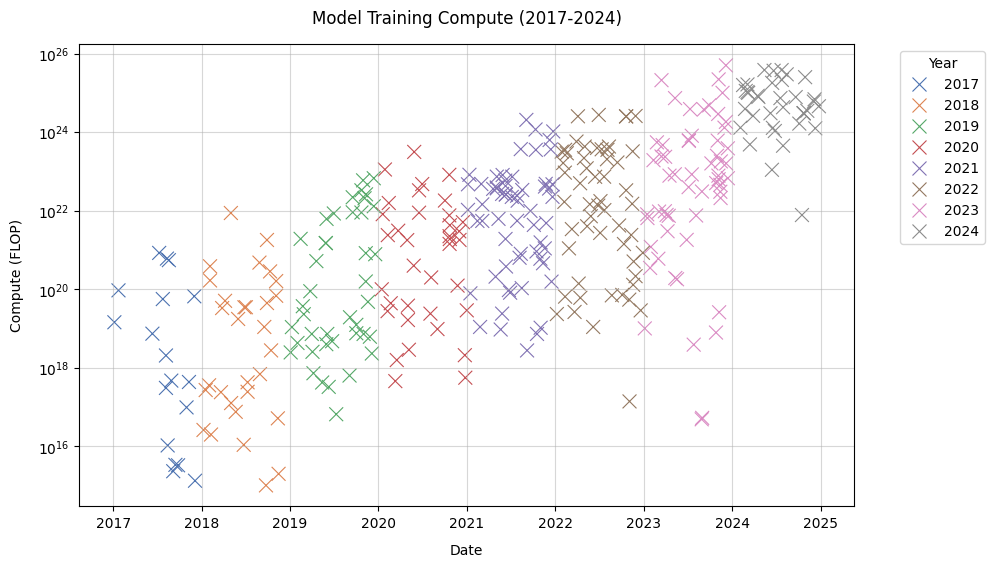}}
\caption{Historical distribution of models over training compute - scatter plot.}
\label{figure:historical-distribution-scatter}
\end{figure}

\begin{figure}[h]
\centering
\fbox{\includegraphics[width=0.5\textwidth]{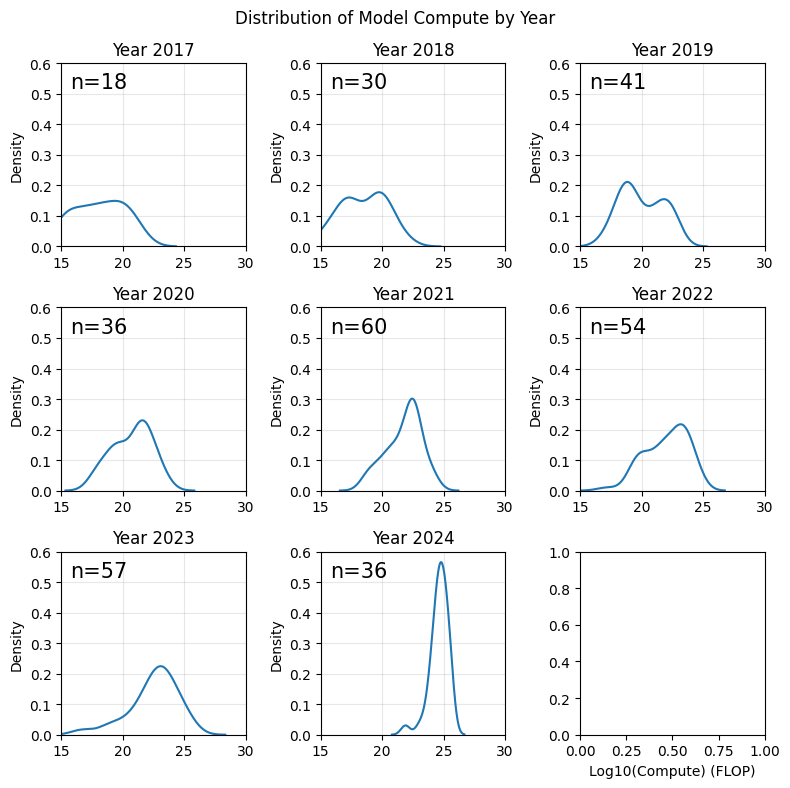}}
\caption{Historical distribution of models over training compute - KDEs}
\label{figure:historical-distribution-kde}
\end{figure}

The dataset we use for this analysis contains records of models and their estimated training compute dating back to the 1950s. We filter out all models that were released before 2017 as this corresponds to the era before the release of the Transformer architecture at the core of most frontier models today. Additionally, we do not use 2024 data to fit our model. Observing the model distribution in Figure \ref{figure:historical-distribution-scatter}, 2024 data appears incomplete relative to the previous years, especially towards the lower end of the distribution (i.e: there is a notable absence of models on the lower end of compute usage). The deviation of 2024 data from previous years can also be seen in the kernel density estimates of Figure \ref{figure:historical-distribution-kde}.
\clearpage

\section{Compute allocations for 2017-2019}
\label{appendix-b-compute-allocations-for-2017-2019}

\begin{figure}[h]
\centering
\fbox{\includegraphics[width=0.8\textwidth]{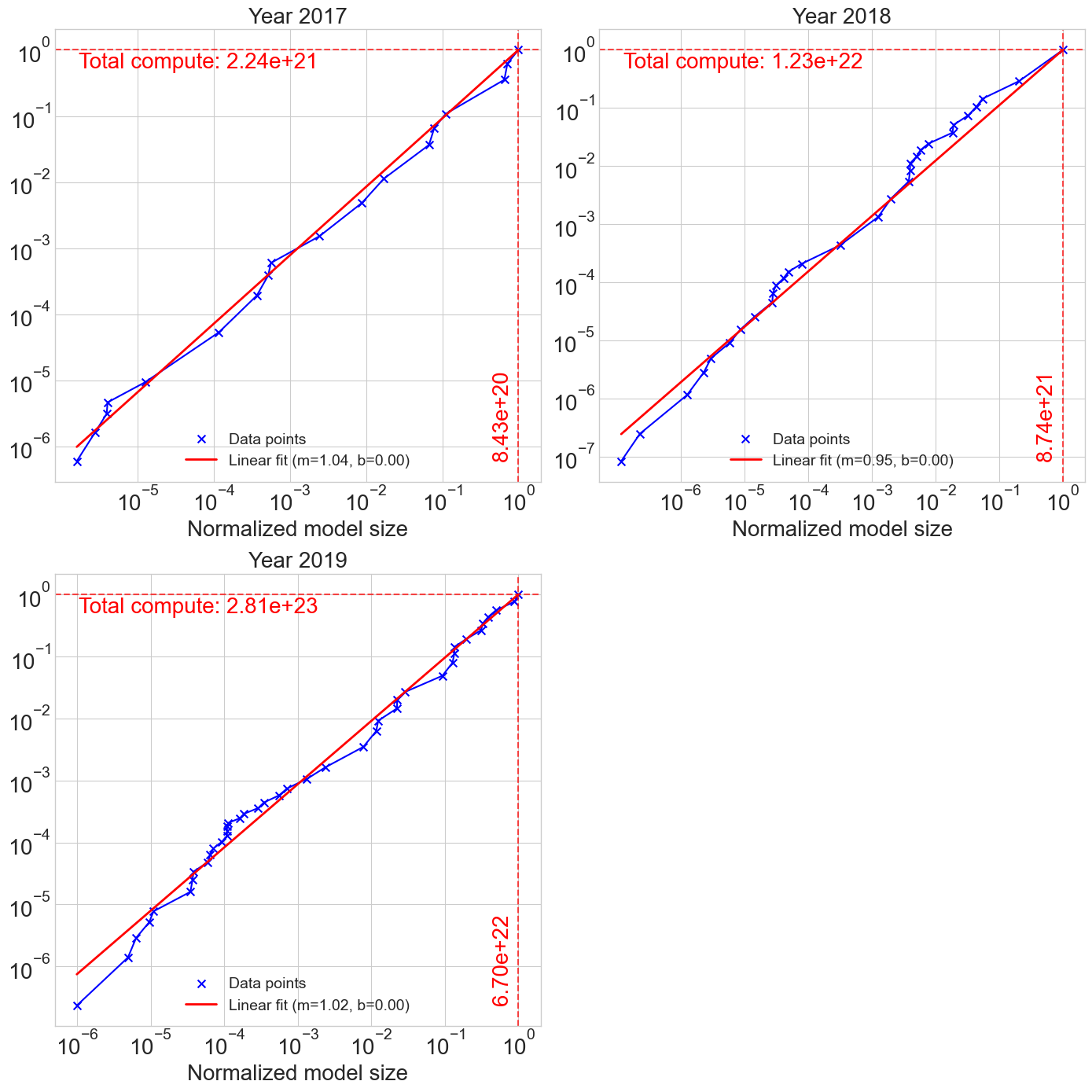}}
\caption{Linear fits to cumulative allocation plots for years 2017-2019.}
\end{figure}
\clearpage

\section{Constraints and interpretation of the linear fits}
\label{appendix-c-constraints-and-interpretation-of-the-linear-fits}

This Appendix discusses the constraints on the linear fit to the compute allocation trends, and the interpretation of the allocation gradient parameter.

Observing historical data we see that the relationship between normalized model size (normalized by the largest model trained that year) - $\tilde{m}$ and the fraction of compute spent on models of size $\tilde{m}$ or less (the cumulative distribution function, denoted by $A(\tilde{m})$) is linear in log-space. Mathematically:
\begin{equation}
\log(A(\tilde{m})) = k \cdot \log(\tilde{m}) + b \tag{1}
\end{equation}

Let that largest model trained in a given year be $m_{\text{max}}$, then $\tilde{m}_{\text{max}} = 1$. Models of size $m_{\text{max}}$ or smaller (i.e., all models) take up all compute spending that year, therefore $A(\tilde{m}_{\text{max}} = 1) = 1$. Enforcing this constraint on equation 1 means that $b = 0$ - and so equation (1) reduces to $A(\tilde{m}) = \tilde{m}^k$.

The parameter $k$ determines how compute is allocated across models of different scales. To see this, let us first denote $a(m_1, m_2)$ as the amount of compute that is allocated to models in the range $[m_1, m_2)$. Consider also three sizes of models - $m^*$, $10m^*$, and $100m^*$. The compute allocated to models in the range $[m^*, 10m^*]$ is $a(m^*, 10m^*) = A(10m^*) - A(m^*) = (10^k - 1)m^{k}$ using equation (1). The compute allocated to models in the range $[10m^*, 100m^*)$ is $a(10m^*, 100m^*) = A(100m^*) - A(10m^*) = 10^k(10^k - 1)m^{k}$ after some simplification. Therefore, the relationship between $a(m^*, 10m^*)$ and $a(10m^*, 100m^*)$ is simply:
\begin{equation}
a(10m^*, 100m^*) = 10^k \cdot a(m^*, 10m^*) \tag{2}
\end{equation}

In other words, scaling up model size by a factor of 10 leads to a factor of $10^k$ increase in compute allocated to models of this size. $k = 1$ means that these larger models get 10 times as much compute as their smaller counterparts. $k > 1$ means that they get a factor greater than 10, and $k < 1$ leads to a factor less than 10.
\clearpage

%% file: texs/9_appendix_d_f.tex
\section{Historical values of allocation gradient \& allocations for varying gradient}
\label{appendix-d-historical-values-of-allocation-gradients}

\begin{figure}[h]
\centering
\fbox{\includegraphics[width=0.8\textwidth]{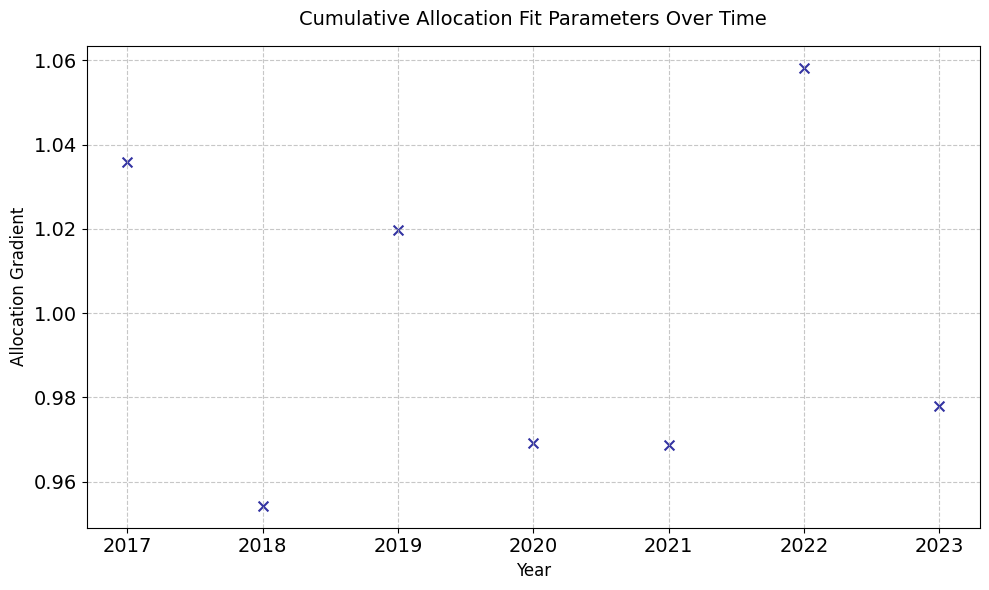}}
\caption{Allocation gradient parameter for 2017-2023.}
\end{figure}

Historical values of the gradient of the compute allocation fits ($k$). Informed by this, our model draws $k$ uniformly from the range [0.9,1.1].
\\

\begin{table}[H]
\centering
\renewcommand{\arraystretch}{1.3}
\small
\resizebox{\textwidth}{!}{
\begin{tabular}{c>{\centering\arraybackslash}c>{\centering\arraybackslash}c>{\centering\arraybackslash}c>{\centering\arraybackslash}c>{\centering\arraybackslash}c>{\centering\arraybackslash}c>{\centering\arraybackslash}c}
\toprule
\textbf{k} & \textbf{[10$^{-7}$-10$^{-6}$]} & \textbf{[10$^{-6}$-10$^{-5}$]} & \textbf{[10$^{-5}$-10$^{-4}$]} & \textbf{[10$^{-4}$-10$^{-3}$]} & \textbf{[10$^{-3}$-10$^{-2}$]} & \textbf{[10$^{-2}$-10$^{-1}$]} & \textbf{[10$^{-1}$-10$^{0}$]} \\
\midrule
0.5 & 0.068 & 0.22 & 0.68 & 2.2 & 6.8 & 22 & 68 \\
0.75 & 0.0026 & 0.015 & 0.082 & 0.46 & 2.6 & 15 & 82 \\
0.9 & 0.00035 & 0.0028 & 0.022 & 0.17 & 1.4 & 11 & 87 \\
1.0 & 9$\times$10$^{-5}$ & 0.0009 & 0.009 & 0.09 & 0.9 & 9 & 90 \\
1.1 & 2.3$\times$10$^{-5}$ & 0.00029 & 0.0037 & 0.046 & 0.58 & 7.3 & 92 \\
1.25 & 3$\times$10$^{-6}$ & 5.3$\times$10$^{-5}$ & 0.00094 & 0.017 & 0.3 & 5.3 & 94 \\
1.5 & 9.7$\times$10$^{-8}$ & 3.1$\times$10$^{-6}$ & 9.7$\times$10$^{-5}$ & 0.0031 & 0.097 & 3.1 & 97 \\
\bottomrule
\end{tabular}
}
\caption{Compute allocations (\%) for various values of k. Each column shows allocation percentages for model size ranges, with models defined with respect to the largest training run. To illustrate: for the k=1.1 case, 92\% of compute is allocated to models within an OOM of the largest model, 7.3\% is allocated to models within 1 and 2 OOMs of the largest model, 0.58\% to models within 2  and 3 OOMs of the largest model, etc.}
\label{tab:compute-allocations-k-values}
\end{table}

\clearpage

\section{Results under uniform sampling of LMS parameter}
\label{appendix-results-uniform-sampling}

\begin{figure}[h]
\centering
\fbox{\includegraphics[width=0.9\textwidth]{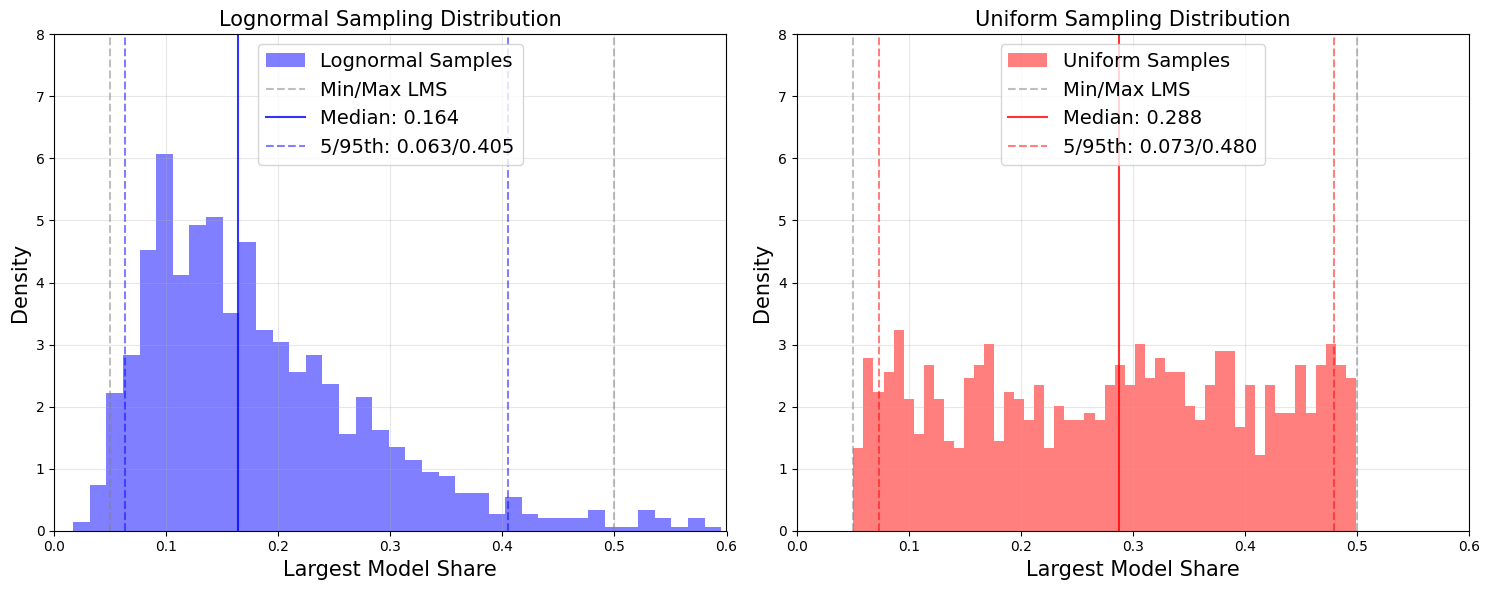}}
\caption{Left: Lognormal LMS distribution with bounds [0.05,0.5] (any value drawn from outside these bounds is resampled) and Right: Uniform LMS distribution. Medians, 5th and 95th percentile LMS indicated.}
\label{figure:lognormal_uniform_LMS_distribution}
\end{figure}

Section \ref{section:lms_modelling_assumptions} discusses the choice to model the largest model share (LMS) parameter for the years 2025-2028  with a lognormal distribution (recall that the 2024 LMS is set such that the largest model is the size of GPT-4o), rather than a uniform distribution. In this appendix, Figure \ref{figure:lognormal_uniform_LMS_distribution} shows the difference between these choices for the LMS sampling distributions, and Table \ref{tab:uniform-LMS-predictions} shows the threshold count results when the LMS parameter is sampled from the uniform distribution instead. In Table \ref{tab:uniform-LMS-predictions}, notably less aggressive medians are observed which results from the median of the LMS distribution being larger than for the lognormal case (for reasons on why larger LMS values lead to less aggressive medians see Appendix \ref{f.2-influence-of-lms-on-projections}).

\begin{table}[htbp]
\centering
\renewcommand{\arraystretch}{1.3}
\begin{tabular}{>{\raggedright\arraybackslash}p{2.2cm}>{\centering\arraybackslash}p{2.2cm}>{\centering\arraybackslash}p{2.2cm}>{\centering\arraybackslash}p{2.2cm}>{\centering\arraybackslash}p{2.2cm}>{\centering\arraybackslash}p{2.2cm}}
\toprule
\textbf{Threshold (FLOP)} & \textbf{2024} & \textbf{2025} & \textbf{2026} & \textbf{2027} & \textbf{2028} \\
\midrule
$\boldsymbol{>}$ \textbf{10$^{25}$} & \mbox{[19, 23, 28]} & \mbox{[31, 42, 67]} & \mbox{[47, 67, 116]} & \mbox{[65, 97, 185]} & \mbox{[88, 137, 287]} \\
$\boldsymbol{>}$ \textbf{10$^{26}$} & \mbox{[0, 0, 0]} & \mbox{[4, 7, 10]} & \mbox{[12, 21, 36]} & \mbox{[23, 40, 76]} & \mbox{[39, 68, 142]} \\
$\boldsymbol{>}$ \textbf{10$^{27}$} & \mbox{[0, 0, 0]} & \mbox{[0, 0, 0]} & \mbox{[0, 3, 5]} & \mbox{[3, 11, 19]} & \mbox{[12, 26, 51]} \\
$\boldsymbol{>}$ \textbf{10$^{28}$} & \mbox{[0, 0, 0]} & \mbox{[0, 0, 0]} & \mbox{[0, 0, 0]} & \mbox{[0, 0, 1]} & \mbox{[0, 4, 9]} \\
$\boldsymbol{>}$ \textbf{10$^{29}$} & \mbox{[0, 0, 0]} & \mbox{[0, 0, 0]} & \mbox{[0, 0, 0]} & \mbox{[0, 0, 0]} & \mbox{[0, 0, 0]} \\
\bottomrule
\end{tabular}
\caption{Results for absolute compute thresholds when the LMS parameter is sampled uniformly from the range [0.05,0.5].}
\label{tab:uniform-LMS-predictions}
\end{table}

\section{LMS parameter and influence of LMS on projections}

\subsection{Historical LMS}
\label{f.1-historical-lms}

\begin{figure}[H]
\centering
\fbox{\includegraphics[width=0.75\textwidth]{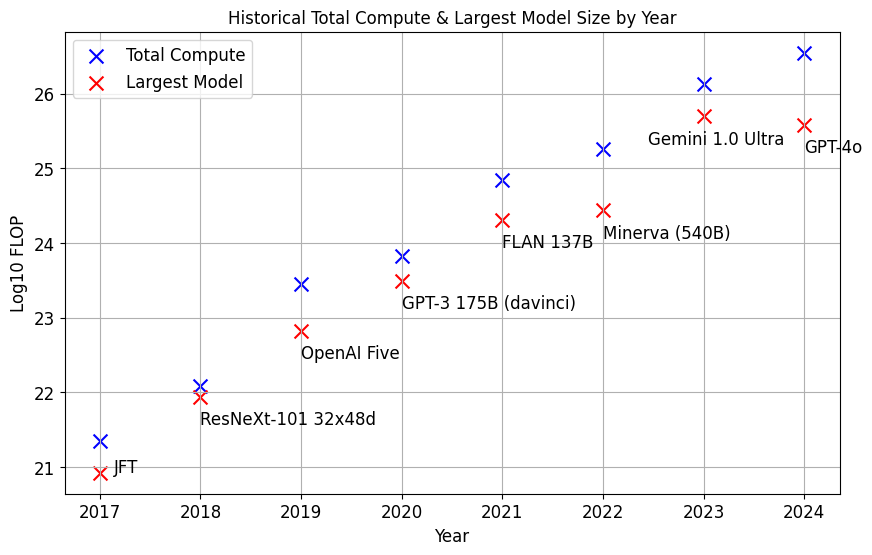}}
\caption{Historical data for the largest model and total training compute.}
\end{figure}

\begin{figure}[H]
\centering
\fbox{\includegraphics[width=0.75\textwidth]{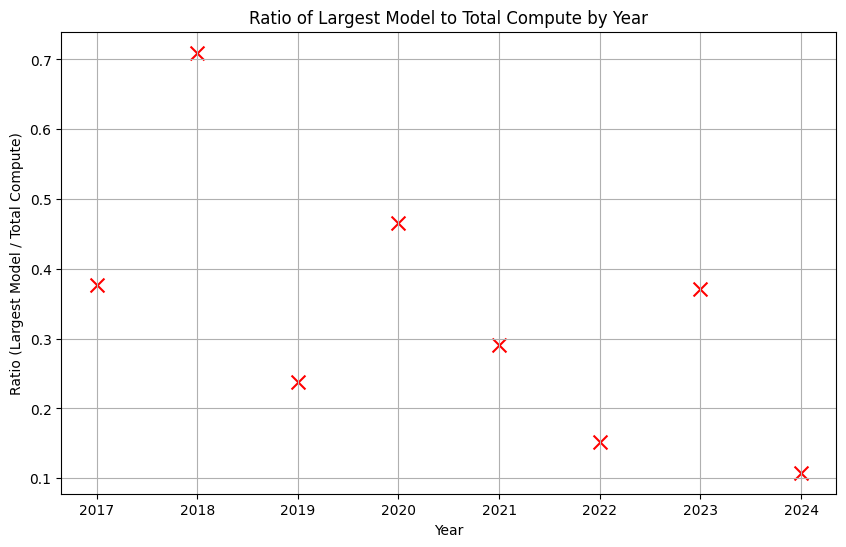}}
\caption{Historical data for the largest model share (LMS) parameter}
\end{figure}

Historical values of the LMS parameter and the total compute and the largest model each year that the LMS is derived from. The AlphaGo family of models have been removed from the dataset on the basis of being outliers, as done in similar analyses \cite{sevilla2022compute}. We fit the model on data from 2017-2023 and also discount the LMS for 2018 as it appears to be an outlier. 

Our predictions sample the LMS uniformly from the range [0.05,0.5]. The upper bound is chosen to accommodate GPT-3 davinci accounting for $\sim$46\% of training compute in 2020. The lower bound is chosen with the 2022 value of 0.15 in mind, however we incorporate a wide range underneath this value due to the strong influence of low LMS values on the model's predictions (see Appendix F.2 below).

\subsection{Influence of LMS on projections}
\label{f.2-influence-of-lms-on-projections}

When experimenting with the model, we find that the \textbf{number} of models that exceed the compute thresholds grows very large as the LMS parameter tends to 0. This appendix explores in further detail why this is the case with a toy example.

Consider the following setup. We have $10^{30}$ FLOP of compute allocated to training models in a given year. The assignment of compute amongst model sizes is shown in the table below. (Note that we assume that increasing the model size by a factor of 10 leads to 10x as much allocated compute - corresponding to k=1 for the allocation gradient.)

\begin{table}[H]
\centering
\renewcommand{\arraystretch}{1.2}
\begin{tabular}{>{\raggedright\arraybackslash}p{3.2cm}>{\centering\arraybackslash}p{2cm}>{\centering\arraybackslash}p{2cm}>{\centering\arraybackslash}p{2cm}>{\centering\arraybackslash}p{2cm}}
\toprule
\textbf{Model size} & \textbf{Within 3-4 OOM} & \textbf{Within 2-3 OOM} & \textbf{Within 1-2 OOM} & \textbf{Within 1 OOM} \\
\midrule
Fractional allocation & 0.009\% & 0.9\% & 9\% & 90\% \\
Allocation (FLOP) & $9\times10^{26}$ & $9\times10^{27}$ & $9\times10^{28}$ & $9\times10^{29}$ \\
\bottomrule
\end{tabular}
\caption{Compute allocation across model size categories, where model size is given relative to the largest training run that year.}
\label{tab:compute-allocation}
\end{table}

Now let us consider two cases: one in which the LMS = 0.05, and another in which the LMS = 0.50. In scenario 1, the largest model trained that year is $0.05\times10^{30}$ = $5\times10^{28}$ FLOP, and in scenario 2, the largest model is $0.5\times10^{30}$ = $5\times10^{29}$ FLOP. Now consider the number of models that can be drawn from each category. To approximate this we find the average model size of each category\footnote{This is done by taking the geometric mean of the model category bounds. We take the geometric mean instead of the arithmetic mean as the bounds are given in log space.} - $m_{avg}$ - and find how many times $m_{avg}$ can be sampled from the compute allocations given in row three of the table above. This is show in the tables below.

In the case where the LMS=0.05, approximately 56 models of size $m_{avg}$ can be sampled for each category, compared to the 5 sampled from each category for the case of LMS=0.5. More generally, the number of average-sized models that can be sampled from each category grows inversely with the size of the average model - and the average sized model of a category grows proportionally with the LMS parameter.

\begin{table}[H]
\centering
\renewcommand{\arraystretch}{1.2}
\begin{tabular}{>{\raggedright\arraybackslash}p{3.5cm}>{\centering\arraybackslash}p{2.2cm}>{\centering\arraybackslash}p{2.2cm}>{\centering\arraybackslash}p{2.2cm}>{\centering\arraybackslash}p{2.2cm}>{\centering\arraybackslash}p{1.5cm}}
\toprule
\textbf{Model characteristics} & \textbf{Within 3-4 OOM} & \textbf{Within 2-3 OOM} & \textbf{Within 1-2 OOM} & \textbf{Within 1 OOM} & \textbf{Total} \\
\midrule
Size range (FLOP) & 5$\times$10$^{24}$- 5$\times$10$^{25}$ & 5$\times$10$^{25}$- 5$\times$10$^{26}$ & 5$\times$10$^{26}$- 5$\times$10$^{27}$ & 5$\times$10$^{27}$- 5$\times$10$^{28}$ & — \\
Average size\footnote{Geometric mean} (FLOP) & 1.6$\times$10$^{24}$ & 1.6$\times$10$^{25}$ & 1.6$\times$10$^{26}$ & 1.6$\times$10$^{27}$ & — \\
Allocation (FLOP) & 9$\times$10$^{26}$ & 9$\times$10$^{27}$ & 9$\times$10$^{28}$ & 9$\times$10$^{29}$ & — \\
Number of samples & 56 & 56 & 56 & 56 & 224 \\
\bottomrule
\end{tabular}
\caption{\textbf{Scenario 1 - LMS=0.05}: Model size categories and compute allocation across different orders of magnitude (OOM) relative to the largest model.}
\label{tab:small_lms_outcomes}
\footnotetext{Geometric mean}
\end{table}

\begin{table}[H]
\centering
\renewcommand{\arraystretch}{1.2}
\begin{tabular}{>{\raggedright\arraybackslash}p{3.5cm}>{\centering\arraybackslash}p{2.2cm}>{\centering\arraybackslash}p{2.2cm}>{\centering\arraybackslash}p{2.2cm}>{\centering\arraybackslash}p{2.2cm}>{\centering\arraybackslash}p{1.5cm}}
\toprule
\textbf{Model characteristics} & \textbf{Within 3-4 OOM} & \textbf{Within 2-3 OOM} & \textbf{Within 1-2 OOM} & \textbf{Within 1 OOM} & \textbf{Total} \\
\midrule
Size range (FLOP) & 5$\times$10$^{25}$- 5$\times$10$^{26}$ & 5$\times$10$^{26}$- 5$\times$10$^{27}$ & 5$\times$10$^{27}$- 5$\times$10$^{28}$ & 5$\times$10$^{28}$- 5$\times$10$^{29}$ & — \\
Average size\footnote{Geometric mean} (FLOP) & 1.6$\times$10$^{25}$ & 1.6$\times$10$^{26}$ & 1.6$\times$10$^{27}$ & 1.6$\times$10$^{28}$ & — \\
Allocation (FLOP) & 9$\times$10$^{26}$ & 9$\times$10$^{27}$ & 9$\times$10$^{28}$ & 9$\times$10$^{29}$ & — \\
Number of samples & 5 & 5 & 5 & 5 & 20 \\
\bottomrule
\end{tabular}
\caption{\textbf{Scenario 2 - LMS=0.5}: Model size categories and compute allocation across different orders of magnitude (OOM) relative to the largest model.}
\label{tab:large_lms_outcomes}
\footnotetext{Geometric mean}
\end{table}

\begin{figure}[H]
\centering
\fbox{\includegraphics[width=0.8\textwidth]{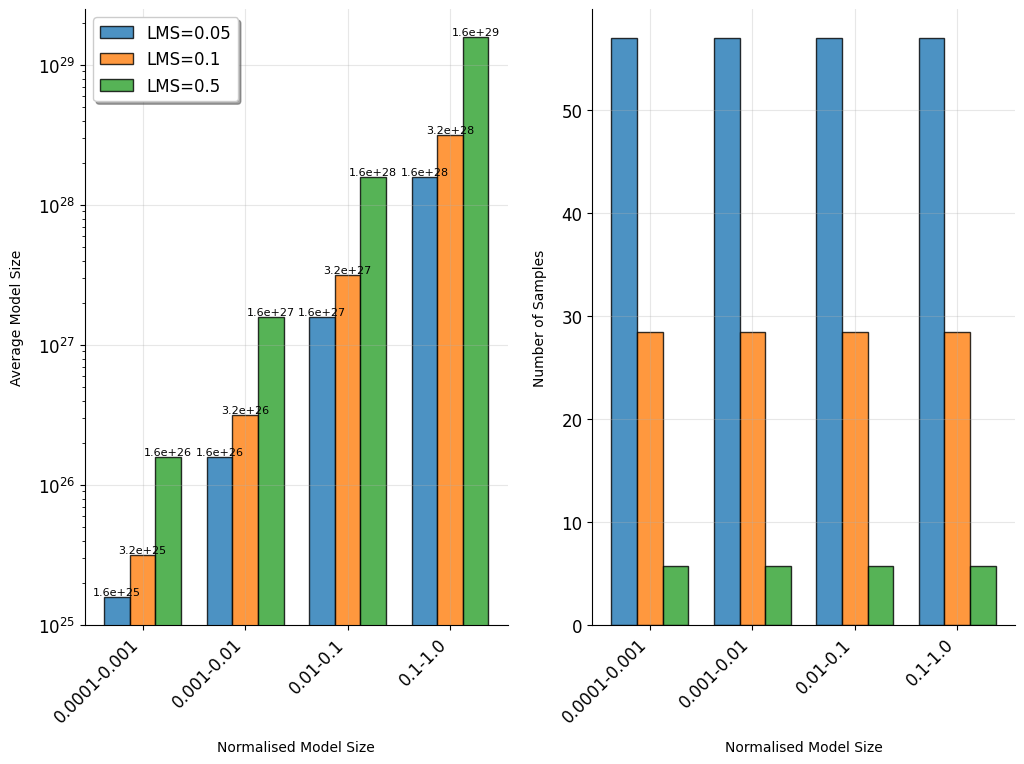}}
\caption{Left: The average-sized model for each category of model size for LMS = 0.05, 0.1, and 0.5. Right: The number of average-sized models that can be drawn from a given compute allocation for LMS = 0.05, 0.1 and 0.5.}
\label{figure:all_lms_outcomes}
\end{figure}

The information in Tables \ref{tab:small_lms_outcomes}, \ref{tab:large_lms_outcomes}  is shown graphically in Figure \ref{tab:small_lms_outcomes}, alongside another configuration where the LMS=0.1. We see that the LMS determines the size of each model category - when the LMS=0.5 the average size of a model in the largest category is $1.6 \times 10^{29}$ FLOP whereas when the LMS = 0.1 and 0.05, the average size of a model in the largest category is $9.5\times10^{28}$ and $1.6\times10^{28}$ respectively. The number of samples that can be drawn given the compute allocations derived above is shown in the right hand side plot - for each category, roughly 50 samples can be drawn when the LMS=0.05, 30 when the LMS=0.1 and 5 when the LMS = 0.5 (the number of samples drawn from each category is constant across model sizes in the plot above because we choose an allocation gradient of k=1.)
\clearpage

%% file: texs/10_appendix_g_i.tex
\section{Results for varying growth rate weightings}
\label{appendix-g-results-for-varying-growth-rate-weightings}

Two sources of evidence inform the growth rates in the baseline scenario - the historical growth rate of training compute stocks and the compute forecast of \cite{compute_forecast}. The baseline predictions gives three times as much weight to the latter (ascribing weights of 0.25 and 0.75 to each respective growth rate), but this is a subjective choice. This appendix shows predictions for other choices of growth rate weightings - namely weightings of (0.1,0.9), (0.33, 0.66) and (0.5, 0.5). As expected, more greater medians estimates are observed when assigning more weight to the larger historical growth rate. 

\begin{table}[H]
\centering
\renewcommand{\arraystretch}{1.3}
\begin{tabular}{>{\raggedright\arraybackslash}p{2.2cm}>{\centering\arraybackslash}p{2.2cm}>{\centering\arraybackslash}p{2.2cm}>{\centering\arraybackslash}p{2.2cm}>{\centering\arraybackslash}p{2.2cm}>{\centering\arraybackslash}p{2.2cm}}
\toprule
\textbf{Threshold (FLOP)} & \textbf{2024} & \textbf{2025} & \textbf{2026} & \textbf{2027} & \textbf{2028} \\
\midrule
$\boldsymbol{>}$ \textbf{10$^{25}$} & \mbox{[15, 20, 24]} & \mbox{[27, 41, 57]} & \mbox{[45, 71, 109]} & \mbox{[63, 106, 179]} & \mbox{[88, 150, 268]} \\
$\boldsymbol{>}$ \textbf{10$^{26}$} & \mbox{[0, 0, 0]} & \mbox{[0, 5, 8]} & \mbox{[8, 18, 31]} & \mbox{[18, 36, 68]} & \mbox{[34, 64, 123]} \\
$\boldsymbol{>}$ \textbf{10$^{27}$} & \mbox{[0, 0, 0]} & \mbox{[0, 0, 0]} & \mbox{[0, 0, 3]} & \mbox{[0, 4, 13]} & \mbox{[2, 15, 37]} \\
$\boldsymbol{>}$ \textbf{10$^{28}$} & \mbox{[0, 0, 0]} & \mbox{[0, 0, 0]} & \mbox{[0, 0, 0]} & \mbox{[0, 0, 0]} & \mbox{[0, 0, 4]} \\
$\boldsymbol{>}$ \textbf{10$^{29}$} & \mbox{[0, 0, 0]} & \mbox{[0, 0, 0]} & \mbox{[0, 0, 0]} & \mbox{[0, 0, 0]} & \mbox{[0, 0, 0]} \\
\bottomrule
\end{tabular}
\caption{Results for absolute thresholds with growth weighting of (0.9,0.1) between historical growth rate (6.3x) and forecasted growth rate (3.4x).}
\label{tab:growth-rate-1}
\end{table}

\begin{table}[H]
\centering
\renewcommand{\arraystretch}{1.3}
\begin{tabular}{>{\raggedright\arraybackslash}p{2.2cm}>{\centering\arraybackslash}p{2.2cm}>{\centering\arraybackslash}p{2.2cm}>{\centering\arraybackslash}p{2.2cm}>{\centering\arraybackslash}p{2.2cm}>{\centering\arraybackslash}p{2.2cm}}
\toprule
\textbf{Threshold (FLOP)} & \textbf{2024} & \textbf{2025} & \textbf{2026} & \textbf{2027} & \textbf{2028} \\
\midrule
$\boldsymbol{>}$ \textbf{10$^{25}$} & \mbox{[19, 24, 29]} & \mbox{[33, 48, 69]} & \mbox{[54, 84, 128]} & \mbox{[78, 129, 207]} & \mbox{[106, 178, 305]} \\
$\boldsymbol{>}$ \textbf{10$^{26}$} & \mbox{[0, 0, 0]} & \mbox{[5, 8, 11]} & \mbox{[16, 26, 38]} & \mbox{[30, 53, 87]} & \mbox{[50, 87, 152]} \\
$\boldsymbol{>}$ \textbf{10$^{27}$} & \mbox{[0, 0, 0]} & \mbox{[0, 0, 0]} & \mbox{[0, 2, 5]} & \mbox{[3, 11, 25]} & \mbox{[13, 30, 57]} \\
$\boldsymbol{>}$ \textbf{10$^{28}$} & \mbox{[0, 0, 0]} & \mbox{[0, 0, 0]} & \mbox{[0, 0, 0]} & \mbox{[0, 0, 1]} & \mbox{[0, 4, 9]} \\
$\boldsymbol{>}$ \textbf{10$^{29}$} & \mbox{[0, 0, 0]} & \mbox{[0, 0, 0]} & \mbox{[0, 0, 0]} & \mbox{[0, 0, 0]} & \mbox{[0, 0, 0]} \\
\bottomrule
\end{tabular}
\caption{Results for absolute thresholds with growth weighting of (0.33,0.66) between historical growth rate (6.3x) and forecasted growth rate (3.4x).}
\label{tab:growth-rate-2}
\end{table}

\begin{table}[H]
\centering
\renewcommand{\arraystretch}{1.3}
\begin{tabular}{>{\raggedright\arraybackslash}p{2.2cm}>{\centering\arraybackslash}p{2.2cm}>{\centering\arraybackslash}p{2.2cm}>{\centering\arraybackslash}p{2.2cm}>{\centering\arraybackslash}p{2.2cm}>{\centering\arraybackslash}p{2.2cm}}
\toprule
\textbf{Threshold (FLOP)} & \textbf{2024} & \textbf{2025} & \textbf{2026} & \textbf{2027} & \textbf{2028} \\
\midrule
$\boldsymbol{>}$ \textbf{10$^{25}$} & \mbox{[22, 26, 31]} & \mbox{[37, 52, 78]} & \mbox{[59, 90, 146]} & \mbox{[83, 136, 239]} & \mbox{[115, 195, 363]} \\
$\boldsymbol{>}$ \textbf{10$^{26}$} & \mbox{[0, 0, 0]} & \mbox{[6, 9, 14]} & \mbox{[19, 31, 48]} & \mbox{[34, 59, 107]} & \mbox{[58, 101, 189]} \\
$\boldsymbol{>}$ \textbf{10$^{27}$} & \mbox{[0, 0, 0]} & \mbox{[0, 0, 0]} & \mbox{[0, 4, 7]} & \mbox{[7, 17, 30]} & \mbox{[21, 41, 77]} \\
$\boldsymbol{>}$ \textbf{10$^{28}$} & \mbox{[0, 0, 0]} & \mbox{[0, 0, 0]} & \mbox{[0, 0, 0]} & \mbox{[0, 0, 2]} & \mbox{[0, 8, 17]} \\
$\boldsymbol{>}$ \textbf{10$^{29}$} & \mbox{[0, 0, 0]} & \mbox{[0, 0, 0]} & \mbox{[0, 0, 0]} & \mbox{[0, 0, 0]} & \mbox{[0, 0, 0]} \\
\bottomrule
\end{tabular}
\caption{Results for absolute thresholds with growth weighting of (0.5,0.5) between historical growth rate (6.3x) and forecasted growth rate (3.4x).}
\label{tab:growth-rate-3}
\end{table}
\newpage

\section{Results for alternate training compute allocations}
\label{appendix-h---results-for-alternate-training-compute-allocations}

Section \ref{allocating-compute-between-training-inference-and-other-workloads} highlighted that the publicly available information on the allocation of compute between training, inference and other workloads was conflicting. Our baseline model uses a slightly adjusted version of the allocation in \cite{compute_forecast}. However another source for the compute allocations is the Epoch GATE model \cite{gate_model} - see the training-inference split graph. Note that the GATE model forecasts the allocation of effective compute - which is the physical compute stock multiplied by algorithmic progress. However in the near future (e.g: 2025-2028), these two quantities do not substantially differ. The allocations for this model are shown below.

\begin{table}[H]
\centering
\renewcommand{\arraystretch}{1.3}
\begin{tabular}{>{\centering\arraybackslash}p{2.5cm}>{\centering\arraybackslash}p{6cm}}
\toprule
\textbf{Year} & \textbf{Approximate training compute allocations (GATE)} \\
\midrule
2024 & 90\% \\
2025 & 90\% \\
2026 & 70\% \\
2027 & 70\% \\
2028 & 70\% \\
\bottomrule
\end{tabular}
\caption{GATE model compute allocations for training, 2025-2028}
\label{tab:gate-allocations-by-year}
\end{table}

It's clear that these forecasts differ substantially to those in AI 2027. One partial explanation could be that the GATE forecasts incorporate compute used for experiments into the training share, however it is out of scope of this article to explore and resolve these discrepancies. Instead, Table \ref{tab:gate-allocation-results} shows the predictions of the model for the absolute compute thresholds when training compute allocations are set as above.

\begin{table}[htbp]
\centering
\renewcommand{\arraystretch}{1.3}
\begin{tabular}{>{\raggedright\arraybackslash}p{2.2cm}>{\centering\arraybackslash}p{2.2cm}>{\centering\arraybackslash}p{2.2cm}>{\centering\arraybackslash}p{2.2cm}>{\centering\arraybackslash}p{2.2cm}>{\centering\arraybackslash}p{2.2cm}}
\toprule
\textbf{Threshold (FLOP)} & \textbf{2024} & \textbf{2025} & \textbf{2026} & \textbf{2027} & \textbf{2028} \\
\midrule
$\boldsymbol{>}$ \textbf{10$^{25}$} & \mbox{[36, 46, 54]} & \mbox{[52, 74, 99]} & \mbox{[71, 114, 172]} & \mbox{[97, 160, 265]} & \mbox{[132, 221, 380]} \\
$\boldsymbol{>}$ \textbf{10$^{26}$} & \mbox{[0, 0, 0]} & \mbox{[8, 13, 18]} & \mbox{[19, 34, 52]} & \mbox{[36, 63, 106]} & \mbox{[59, 105, 183]} \\
$\boldsymbol{>}$ \textbf{10$^{27}$} & \mbox{[0, 0, 0]} & \mbox{[0, 0, 2]} & \mbox{[0, 4, 11]} & \mbox{[7, 17, 34]} & \mbox{[20, 40, 77]} \\
$\boldsymbol{>}$ \textbf{10$^{28}$} & \mbox{[0, 0, 0]} & \mbox{[0, 0, 0]} & \mbox{[0, 0, 0]} & \mbox{[0, 0, 4]} & \mbox{[0, 7, 20]} \\
$\boldsymbol{>}$ \textbf{10$^{29}$} & \mbox{[0, 0, 0]} & \mbox{[0, 0, 0]} & \mbox{[0, 0, 0]} & \mbox{[0, 0, 0]} & \mbox{[0, 0, 0]} \\
\bottomrule
\end{tabular}
\caption{Results for absolute compute thresholds with GATE model allocations between training and inference compute.}
\label{tab:gate-allocation-results}
\end{table}

\section{Results for varying allocation gradients}
\label{section:appendix-varying-allocation-gradient}

\begin{table}[H]
\centering
\renewcommand{\arraystretch}{1.3}
\small
\resizebox{\textwidth}{!}{
\begin{tabular}{c>{\centering\arraybackslash}c>{\centering\arraybackslash}c>{\centering\arraybackslash}c>{\centering\arraybackslash}c>{\centering\arraybackslash}c>{\centering\arraybackslash}c>{\centering\arraybackslash}c}
\toprule
\textbf{k} & \textbf{10$^{-7}$--10$^{-6}$} & \textbf{10$^{-6}$--10$^{-5}$} & \textbf{10$^{-5}$--10$^{-4}$} & \textbf{10$^{-4}$--10$^{-3}$} & \textbf{10$^{-3}$--10$^{-2}$} & \textbf{10$^{-2}$--10$^{-1}$} & \textbf{10$^{-1}$--10$^{0}$} \\
\midrule
0.5 & 0.068 & 0.22 & 0.68 & 2.2 & 6.8 & 22 & 68 \\
0.6 & 0.019 & 0.075 & 0.3 & 1.2 & 4.7 & 19 & 75 \\
0.7 & 0.0051 & 0.025 & 0.13 & 0.64 & 3.2 & 16 & 80 \\
0.8 & 0.0013 & 0.0084 & 0.053 & 0.34 & 2.1 & 13 & 84 \\
0.9 & 0.00035 & 0.0028 & 0.022 & 0.17 & 1.4 & 11 & 87 \\
1.0 & 0.00009 & 0.0009 & 0.009 & 0.09 & 0.9 & 9 & 90 \\
\bottomrule
\end{tabular}
}
\caption{Compute allocations (\%) for various values of k $<$ 1. Each column shows allocation percentages for model size ranges relative to the largest model trained. For example, the second column from the right shows the compute allocations for models within 1 and 2 OOMs of the largest model trained.}
\label{tab:compute-allocations-k-values}
\end{table}

Our baseline scenario samples the allocation gradient uniformly from the range [0.9, 1.1]. The median prediction in this scenario will therefore follow a compute allocation across model sizes as shown in the k=1 scenario in the table above. This modeling choice is made from observations of the allocation plots for the Notable models released in the years 2017-2023 (Appendix \ref{appendix-d-historical-values-of-allocation-gradients}). 

However Section \ref{parameter-uncertainty-section} discusses the limitations of the Notable Models database upon which these trends are based. One potential way to account for the Notable Models selection effect is to allocate more compute to smaller models relative to their larger counterparts. This can be seen in the table above where the k=0.5 case allocates $\sim$68\% of compute that year to the largest model category, whereas the k=1.0 case allocates 90\% of compute. More generally, increasing model size by 10x leads to a 10$^k$ times increase in compute allocated, as shown in Appendix \ref{appendix-c-constraints-and-interpretation-of-the-linear-fits}.

This appendix presents model predictions for allocation gradients that allocate relatively more compute to smaller model sizes. Specifically, Table \ref{tab:allocation-gradient-0.7-0.9} presents the results of the model when the allocation gradient (k) is sampled from the range [0.7,0.9] (corresponding to a median scenario in which k = 0.8), and Table \ref{tab:allocation-gradient-0.5-0.7} presents the results of the model when the allocation gradient is sampled from the range [0.5,0.7] (corresponding to a median scenario in which k = 0.6). Notably more aggressive medians can be observed in the later years of the projection compared to the baseline - this is because these scenarios allocate relatively more compute to smaller models than the baseline, and in the 2027 and 2028, 10$^{25}$ and 10$^{26}$ FLOP models are multiple orders of magnitude away from the frontier models. 

\begin{table}[H]
\centering
\renewcommand{\arraystretch}{1.3}
\begin{tabular}{>{\raggedright\arraybackslash}p{2.2cm}>{\centering\arraybackslash}p{2.2cm}>{\centering\arraybackslash}p{2.2cm}>{\centering\arraybackslash}p{2.2cm}>{\centering\arraybackslash}p{2.2cm}>{\centering\arraybackslash}p{2.2cm}}
\toprule
\textbf{Threshold (FLOP)} & \textbf{2024} & \textbf{2025} & \textbf{2026} & \textbf{2027} & \textbf{2028} \\
\midrule
$\boldsymbol{>}$ \textbf{10$^{25}$} & \mbox{[36, 46, 54]} & \mbox{[52, 74, 99]} & \mbox{[71, 114, 172]} & \mbox{[97, 160, 265]} & \mbox{[132, 221, 380]} \\
$\boldsymbol{>}$ \textbf{10$^{26}$} & \mbox{[0, 0, 0]} & \mbox{[8, 13, 18]} & \mbox{[19, 34, 52]} & \mbox{[36, 63, 106]} & \mbox{[59, 105, 183]} \\
$\boldsymbol{>}$ \textbf{10$^{27}$} & \mbox{[0, 0, 0]} & \mbox{[0, 0, 2]} & \mbox{[0, 4, 11]} & \mbox{[7, 17, 34]} & \mbox{[20, 40, 77]} \\
$\boldsymbol{>}$ \textbf{10$^{28}$} & \mbox{[0, 0, 0]} & \mbox{[0, 0, 0]} & \mbox{[0, 0, 0]} & \mbox{[0, 0, 4]} & \mbox{[0, 7, 20]} \\
$\boldsymbol{>}$ \textbf{10$^{29}$} & \mbox{[0, 0, 0]} & \mbox{[0, 0, 0]} & \mbox{[0, 0, 0]} & \mbox{[0, 0, 0]} & \mbox{[0, 0, 0]} \\
\bottomrule
\end{tabular}
\caption{Results for absolute compute thresholds when the allocation gradient is sampled uniformly from the range [0.7,0.9].}
\label{tab:allocation-gradient-0.7-0.9}
\end{table}

\begin{table}[H]
\centering
\renewcommand{\arraystretch}{1.3}
\begin{tabular}{>{\raggedright\arraybackslash}p{2.2cm}>{\centering\arraybackslash}p{2.2cm}>{\centering\arraybackslash}p{2.2cm}>{\centering\arraybackslash}p{2.2cm}>{\centering\arraybackslash}p{2.2cm}>{\centering\arraybackslash}p{2.2cm}}
\toprule
\textbf{Threshold (FLOP)} & \textbf{2024} & \textbf{2025} & \textbf{2026} & \textbf{2027} & \textbf{2028} \\
\midrule
$\boldsymbol{>}$ \textbf{10$^{25}$} & \mbox{[15, 19, 23]} & \mbox{[36, 49, 67]} & \mbox{[70, 106, 155]} & \mbox{[116, 199, 314]} & \mbox{[205, 359, 637]} \\
$\boldsymbol{>}$ \textbf{10$^{26}$} & \mbox{[0, 0, 0]} & \mbox{[2, 5, 9]} & \mbox{[12, 21, 34]} & \mbox{[29, 55, 88]} & \mbox{[63, 113, 196]} \\
$\boldsymbol{>}$ \textbf{10$^{27}$} & \mbox{[0, 0, 0]} & \mbox{[0, 0, 0]} & \mbox{[0, 1, 4]} & \mbox{[3, 7, 16]} & \mbox{[12, 26, 52]} \\
$\boldsymbol{>}$ \textbf{10$^{28}$} & \mbox{[0, 0, 0]} & \mbox{[0, 0, 0]} & \mbox{[0, 0, 0]} & \mbox{[0, 0, 0]} & \mbox{[0, 1, 6]} \\
$\boldsymbol{>}$ \textbf{10$^{29}$} & \mbox{[0, 0, 0]} & \mbox{[0, 0, 0]} & \mbox{[0, 0, 0]} & \mbox{[0, 0, 0]} & \mbox{[0, 0, 0]} \\
\bottomrule
\end{tabular}
\caption{Results for absolute compute thresholds when the allocation gradient is sampled uniformly from the range [0.5,0.7].}
\label{tab:allocation-gradient-0.5-0.7}
\end{table}
\newpage